\documentclass[letterpaper, 10 pt, journal, twoside]{IEEEtran}

\IEEEoverridecommandlockouts                              

\usepackage{amsmath}
\usepackage{mathtools}
\usepackage{amssymb}
\usepackage{mathabx}        
\usepackage{bbold}          
\usepackage{bm}
\usepackage{soul}
\usepackage{xcolor}

\usepackage{enumitem}
\usepackage{url}
\usepackage{comment}
\usepackage{subfigure}
\usepackage{tabularx}
\usepackage{booktabs}
\usepackage[hidelinks]{hyperref}

\usepackage{tikz}
\usetikzlibrary{calc,angles,positioning,intersections,quotes,decorations.markings}

\newcommand{\RN}{\mathbb{R}} 				

\newcommand{\BR}[1]{\left({ #1 }\right)}	
\newcommand{\CBR}[1]{\left\lbrace #1 \right\rbrace}
\newcommand{\norm}[1]{\left\Vert #1\right\Vert}	
\newcommand{\ABS}[1]{\lvert #1 \rvert}
\newcommand{\PARDIFF}[2]{\ensuremath{\frac{\partial #1}{\partial #2}}}
\newcommand{\DIFF}[2]{\ensuremath{\frac{\text{d}#1}{\text{d}#2}}}

\newcommand{\SPAN}[1]{\textbf{span}\CBR{ #1 }}
\newcommand{\NULL}[1]{\textbf{Null}\BR{ #1 }}

\newcommand{\TRACE}[1]{\textbf{trace}\BR{ #1 }}
\newcommand{\DET}[1]{\textbf{det}\BR{ #1 }}

\newcommand{\DIAG}[1]{\textbf{diag}\BR{ #1 }}

\newcommand{\CARD}[1]{\textbf{card}\BR{ #1 }}
\newtheorem{remark}{Remark}
\newtheorem{theorem}{Theorem}
\newtheorem{lemma}{Lemma}
\newtheorem{proposition}{Proposition}

\newcounter{mytempeqncnt}

\title{\LARGE \bf
Stability Analysis of Gradient-Based Distributed Formation Control with Heterogeneous Sensing Mechanism: Two and Three Robot Case
}
\author{Nelson P.K. Chan$^{\star}$, Bayu Jayawardhana$^{\star}$, and Hector Garcia de Marina$^{\dagger}$
\thanks{
    {The work of N.P.K. Chan \& B. Jayawardhana is supported by the Region of Smart Factories (ROSF) project financed by REP-SNN and by the STW Smart Industry 2016 programme. 
    The work of H.G. de Marina is supported by the grant \emph{Atraccion de Talento} 2019-T2/TIC-13503 from the Government of the Autonomous Community of Madrid.}
    }
\thanks{$^{\star}$N.P.K. Chan and B. Jayawardhana are with 
    Engineering and Technology Institute Groningen, 
    University of Groningen, 
    Nijenborgh 4, 9747AG,
    Groningen, the Netherlands
    (email: {\tt\small n.p.k.chan@rug.nl, b.jayawardhana@rug.nl})
    }%
\thanks{$^{\dagger}$H.G. de Marina is with 
    the Department of Computer Architecture and Automatic Control 
    at the Faculty of Physics, 
    Universidad Complutense de Madrid, 28040, Madrid, Spain
    (email: {\tt\small hgarciad@ucm.es})
    }%
}

\begin{document}

\maketitle
\thispagestyle{empty}
\pagestyle{empty}

\begin{abstract}
This paper focuses on the stability analysis of a formation shape displayed by a team of mobile robots that uses heterogeneous sensing mechanism.
Depending on the convenience and reliability of the local information, each robot utilizes the popular gradient-based control law which, in this paper, is either the distance-based or the bearing-only formation control. 
For the two and three robot case, we show that the use of heterogeneous gradient-based control laws can give rise to an undesired invariant set where a distorted formation shape is moving at a constant velocity.
The (in)stability of such an invariant set is dependent on the specified distance and bearing constraints. 
For the two robot case, we prove almost global stability of the desired equilibrium set while for the three robot case, we guarantee local asymptotic stability for the correct formation shape. 
We also derive conditions for the three robot case in which the undesired invariant set is locally attractive. 
Numerical simulations are presented for illustrating the theoretical results in the three robot case.
\end{abstract}


\section{INTRODUCTION} \label{sec:Introduction}
\IEEEPARstart{F}{or} the past two decades, formation control has been an active research topic within the cooperative control of multi-agent systems. 
Two problems in formation control are the \textit{formation stabilization problem} and the \textit{formation tracking problem}.
In the former, the goal is to design control laws for each agent (in this paper, mobile robot) with the aim to steer the team to display a desired geometric shape in a collective manner.
The latter problem adds to the former the requirement that the whole formation needs to follow a given reference trajectory. 
In the current study, our focus will be on the former problem (also known as \textit{formation shape control}). 

Over the years, a rich body of work has been developed on the problem of realizing a formation shape by a team of mobile robots. 
In these works, the variables for specifying the formation shape are commonly given in terms of a certain set of inter-robot relative positions \cite{Ren2008}, distances \cite{Krick2009, Dorfler2009}, bearings \cite{Zhao2018}, or inner angles \cite{Chen2019}. 
When the formation shape is specified by inter-robot relative positions, then the consensus-based formation control approaches, which are linear, can be used directly. 
For realizing a formation shape using only distance, bearing, or angle constraints, the notion of rigidity \cite{Chen2019, Anderson2008, Zhao2019} for the underlying interconnection topology is required. 
The condition of rigidity describes the motions for the whole formation which preserve the formation shape. 
Among others, gradient-based control laws are a popular approach for realizing the formation shape specified by distance \cite{Sun2016} or bearing constraints \cite{Zhao2018}.
A formation shape can also be realized by a set of mixed constraints. 
In this setting, \cite{Bishop2014} explores the combination of distance and bearing constraints, \cite{Kwon2018} considers distance and angle constraints while in \cite{Kwon2019a}, the combination of distance, bearing and angle constraints is being considered. 
In distance-constrained and also distance-and-angle-constrained formation control, flip and flex ambiguities occur. 
To resolve this problem, signed constraints \cite{Anderson2017, Sugie2018, Kwon2019} have been introduced.

In most of the aforementioned references, the underlying interconnection topology is that of an \textit{undirected} graph. 
This implies that the geometrical constraint between a robot pair, represented by an edge on the graph, is actively maintained by both robots. 
While the use and active maintenance of a common type of constraints (distance, bearing, angle) between two neighboring robots have been proven to guarantee the stability of the desired formation shape, it may not be desirable in practice.
Firstly, the sole use of one type of constraints on all edges implies that all robots must use the same sensing mechanism.
This restriction prevents the integration of other robots carrying different sensor loads into the formation.
Secondly, when multiple types of constraints need to be actively maintained by certain robots, these robots have to be equipped with multiple different sensing mechanisms corresponding to the different constraints defined on the edges connected to them. 
This implies that these robots have to carry extra sensor loads that can be costly and consume additional space, weight and computational load. 
On the other hand, it might be desirable for robots to control only distances or bearings depending on their current situation. 
For instance, the bearing measurement becomes less sensitive when the robots move in a large shape formation, or the robots carrying multiple distance and bearing sensors can control the most relevant constraint depending on the accuracy or reliability of the equipped sensor for a given situation (e.g., far versus near, wide-angle versus small-angle, etc.).
Lastly, if a partial sensor failure in a robot occurs, maintaining the same constraint with its neighbours may no longer be possible. 
For instance, in the case of a partial failure of a LIDAR sensor, which can normally provide relative position information, we may still measure bearing information with non-accurate distance information. 
Distance-based/bearing-based controllers are tolerant to non-accurate bearing/distance measurements \cite{Zhao2019, Bishop2014}. 
In this case, while it is possible to define heterogeneous constraints on the same edge that still define the same shape (e.g., one robot controls relative position while the other one controls bearing), it remains an open problem whether the application of standard local gradient-based control law based on the information available to each robot can still maintain the formation.
Intuitively, the gradient-based control law will direct each robot to the direction that minimizes the local potential function and reaches the desired constraints. 
However, as different types of potential function may be defined for the same edge due to the heterogeneous sensing mechanisms between the two robots, the direction that is taken by each robot may not coincide anymore with the minimization of the combined potential functions.

In this work, we consider the formation stabilization problem in which the desired formation shape is specified by a mixed set of distance and bearing constraints.
Different from \cite{Bishop2014}, we consider the setup where each constraint is actively maintained by \textit{only} one robot, i.e., the underlying interconnection topology is that of a \textit{directed} graph. 
Each robot within the team has the individual task of maintaining a subset of either the distance or bearing constraints. 
For this particular work, we focus on teams consisting of two and three robots in different setups. 
Using standard gradient-based control laws specific to the constraints each robot has to maintain, we analyze the stability property, particularly, the local asymptotic stability of the desired formation shape.
It is of interest to study the applicability of these control laws without modifying their local potential functions to incorporate the different constraints on the edges since it allows us to design distributed control laws for each robot that is completely dependent on the available local information to the robot and is independent of the eventual deployment of the robot in the formation (provided of course that the desired admissible constraints are available to the robot).  

We firstly present some preliminaries and problem formulation in Section \ref{sec:Preliminaries}. 
We start our exposition of the problem by analyzing the stability of a simple two robot case in Section \ref{sec:The-1D1B-Setup}. 
In this section, we show that the deployment of two different gradient-based control laws results in the existence of an invariant set where a distorted formation shape is moving at a constant velocity. 
Despite the existence of such an invariant set, the heterogeneous control laws can still guarantee almost global stability of the desired formation shape; in the current setting, this is a static link.
We further extend the analysis to the three robot case in Sections \ref{sec:The-1D2B-Setup} and \ref{sec:The-1B2D-Setup}.
Section \ref{sec:The-1D2B-Setup} deals with the case when one robot has to maintain distance constraints (w.r.t. the other two robots) and two robots have to maintain bearing constraints. 
Similarly, in Section \ref{sec:The-1B2D-Setup}, we analyze the case in which two robots have to maintain distance constraints and one robot has to maintain bearing constraints. 
Numerical simulations are presented in Section \ref{sec:Numerical-Example} for illustrating the various different final formations in the three robot case.
Finally, conclusions and future work are presented in Section \ref{sec:Conclusions}.
The proofs for some technical results are given in the Appendix.

\vspace{0.5\baselineskip}
\paragraph*{Notation}
The cardinality of a given set $ \mathcal{S} $ is denoted by $ \CARD{\mathcal{S}} $.
For a vector $ x \in \RN^{n} $, $ x^{\top} $ is the transpose and $ \norm{x} = \sqrt{x^{\top}x} $ is the $ 2 $-norm of $ x $. 
The vector $ \mathbb{1}_{n} $ (or $ \mathbb{0}_{n} $) denotes the vector with entries being all $ 1 $s (or $ 0 $s).
The set of all combinations of linearly independent vectors $ v_{1}, \, \dots, \, v_{k} $ is denoted by $ \SPAN{v_{1}, \, \dots, \, v_{k}} $. 
The symbol $ \angle v_{1} $ denotes the counter-clockwise angle from the $ x $-axis of a coordinate frame $ \Sigma $ to the vector $ v_{1} \in \RN^{2} $. 
For a complex number $ z = a + b \, \text{i} $, the numbers $ a, \, b \in \RN $ are the real and imaginary part of $ z $ and $ \text{i}^{2} = -1 $. 
The complex conjugate of $ z $ is $ \widebar{z} = a - b \, \text{i} $.
In polar form, we write $ z = r_{z} \angle \varphi_{z} $, where $ r_{z} = \sqrt{z\widebar{z}} = \sqrt{a^{2} + b^{2}} $ is the modulus and $ \varphi_{z} = \tan^{-1} \BR{\frac{b}{a}} $ is the argument corresponding to $ z $.
Furthermore, $ a = r_{z} \cos \varphi_{z} $, $ b = r_{z} \sin \varphi_{z} $, and $ \widebar{z} = r_{z} \angle \BR{-\varphi_{z}} $.
Let $ y = r_{y} \angle \varphi_{y} $; we have the multiplication $ yz = \BR{r_{y} r_{z}} \angle \BR{\varphi_{y} + \varphi_{z}} $ and the division $ \frac{y}{z} = \BR{\frac{r_{y}}{r_{z}}} \angle \BR{\varphi_{y} - \varphi_{z}} $.
For a matrix $ A \in \RN^{m \times n} $, $ \NULL{A} \subset \RN^{n} $, $ \TRACE{A} $, and $ \DET{A} $ denote the null space, the trace, and the determinant of $ A $, respectively.
The $ n \times n $ identity matrix is denoted by $ I_{n} $ while $ \DIAG{v} $ is the diagonal matrix with entries of vector $ v $ on the main diagonal.
Finally, given matrices $ A \in \RN^{m \times n} $ and $ B \in \RN^{p \times q} $, the Kronecker product is $ A \otimes B \in \RN^{mp \times nq} $, and we denote $ \widetilde{A} = A \otimes I_{d} \in \RN^{md \times nd} $. 


\section{PRELIMINARIES} \label{sec:Preliminaries}
In this section, we provide the background material which will be used later in the paper. 
Furthermore, details on the problem setup are also given. 

\subsection{Graph theory} \label{subsec:Graph-Theory}
\begin{figure}[!tb]
    \centering
    \includegraphics[width=0.95\linewidth]{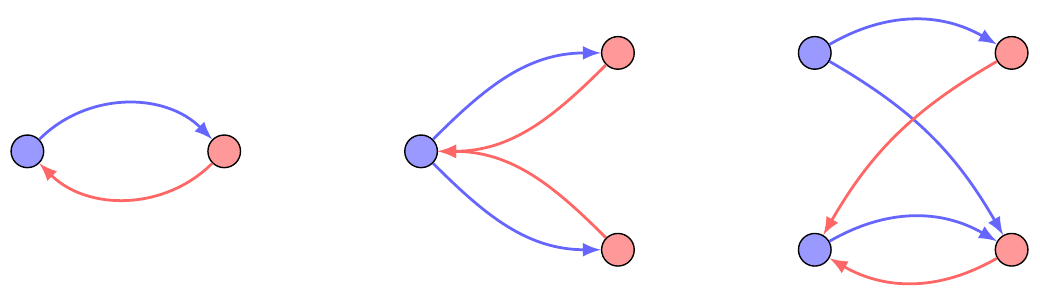}
    \caption{Examples of complete bipartite digraphs for $ n = 2 $ and $ 3 $ vertices and a bipartite digraph for $ n = 4 $ vertices.
    Without loss of generality, \protect\tikz{\protect\draw[fill = blue!40!white] (0, 0) circle [radius = 4pt];} represents an element of $ \mathcal{V}_{1} $ while \protect\tikz{\protect\draw[fill = red!40!white] (0, 0) circle [radius = 4pt];} is a vertex belonging to $ \mathcal{V}_{2} $.
    Correspondingly, the blue arrows are edges belonging to $ \mathcal{V}_{1} \times \mathcal{V}_{2} $ while the red arrows are elements of the edge set $ \mathcal{V}_{2} \times \mathcal{V}_{1} $.
    }
    \label{fig:Graph-Topology}
\end{figure}

A \textit{directed graph} (in short, \textit{digraph}) $ \mathcal{G} $ is a pair $ \BR{\mathcal{V}, \, \mathcal{E}} $, where $ \mathcal{V} = \CBR{ 1, \, 2, \, \dots, \, n} $ is the \textit{vertex} set and $ \mathcal{E} \subseteq \mathcal{V} \times \mathcal{V} $ is the \textit{edge} set.
For $ i, \, j \in \mathcal{V} $, the ordered pair $ \BR{i, \, j} $ represents an edge pointing \textit{from} $ i $ \textit{to} $ j $.
We assume $ \mathcal{G} $ does not have self-loops, i.e., $ \BR{i, \, i} \not\in \mathcal{E} $ for all $ i \in \mathcal{V} $ and $ \CARD{\mathcal{E}} = m $. 
The set of neighbors of vertex $ i $ is denoted by $ \mathcal{N}_{i} = \CBR{ j \in \mathcal{V} \, \rvert \, \BR{i, \, j} \in \mathcal{E}} $.
Associated to $ \mathcal{G} $, we define the incidence matrix $ H \in \CBR{0, \, \pm 1}^{m \times n} $ with entries $ \left[H\right]_{ki} = -1 $ if vertex $ i $ is the tail of edge $ k $, $ \left[H\right]_{ki} = 1 $ if it is the head, and $ \left[H\right]_{ki} = 0 $ otherwise.
Due to its structure, we have $ \SPAN{\mathbb{1}_{n}} \subseteq \NULL{H} $.
The digraph $ \mathcal{G} $ is \textit{bipartite} if the vertex set $ \mathcal{V} $ can be partitioned into two subsets $ \mathcal{V}_{1} $ and $ \mathcal{V}_{2} $ with $ \mathcal{V}_{1} \cap \mathcal{V}_{2} = \emptyset $ and the edge set $ \mathcal{E} \subseteq \BR{\mathcal{V}_{1} \times \mathcal{V}_{2}} \cup \BR{\mathcal{V}_{2} \times \mathcal{V}_{1}} $.
We assume $ \CARD{\mathcal{V}_{1}} = n_{1} $ and $ \CARD{\mathcal{V}_{2}} = n - n_{1} = n_{2} $.
For a \textit{complete bipartite digraph}, the edge set is $ \mathcal{E} = \BR{\mathcal{V}_{1} \times \mathcal{V}_{2}} \cup \BR{\mathcal{V}_{2} \times \mathcal{V}_{1}} $ and $ \CARD{\mathcal{E}} = 2 n_{1} n_{2} $.
Fig. \ref{fig:Graph-Topology} depicts complete bipartite digraphs for $ n = 2 $ and $ 3 $ vertices and a bipartite digraph for $ n = 4 $ vertices.

\subsection{Formations and gradient-based control laws} \label{subsec:FormationGradient-Control-Laws}
We consider a team consisting of $ n $ robots in which \texttt{Ri} is the label assigned to robot $ i $.
The robots are moving in the plane according to the single integrator dynamics, i.e., 
\begin{equation} \label{eq:nR-Robot-Dynamics}
    \begin{aligned}
        \dot{p}_{i} = u_{i}, \quad i \in \CBR{1, \, \dots, \, n}
        ,
    \end{aligned}
\end{equation}
where $ p_{i} \in \RN^{2} $ (a point in the plane) and $ u_{i} \in \RN^{2} $ represent the position of and the control input for \texttt{Ri}, respectively.
For convenience, all spatial variables are given relative to a global coordinate frame $ \Sigma^{\text{g}} $. 
The group dynamics is obtained as $ \dot{p} = u $ with the stacked vectors $ p = {\begin{bmatrix} p_{1}^{\top} & \cdots & p_{n}^{\top} \end{bmatrix}}^{\top} \in \RN^{2n} $ representing the team \textit{configuration} and $ u = {\begin{bmatrix} u_{1}^{\top} & \cdots & u_{n}^{\top} \end{bmatrix}}^{\top} \in \RN^{2n} $ being the collective input. 
The interactions among the robots is described by a fixed graph $ \mathcal{G}\BR{\mathcal{V}, \: \mathcal{E}} $ with $ \mathcal{V} $ being the team of robots and $ \mathcal{E} $ containing the neighboring relationships. 
We can embed the graph $ \mathcal{G} $ into the plane by assigning to each vertex $ i \in \mathcal{V} $, a point $ p_{i} \in \RN^{2} $.
The pair $ \mathcal{F}_{p} = \BR{\mathcal{G}, \: p} $ denotes a \textit{framework} (or equivalently a \textit{formation}) in $ \RN^{2} $.
We assume $ p_{i} \neq p_{j} $ if $ i \neq j $, i.e., two robots cannot be at the same position. 
In addition, we introduce the following notation before providing details on the \textit{distance-based} and \textit{bearing-only} formation control approaches.
For points $ p_{i} $ and $ p_{j} $ of the formation, we define the relative position vector as $ z_{ij} = p_{j} - p_{i} \in \RN^{2} $, the distance as $ d_{ij} = \norm{z_{ij}} \in \RN_{> 0} $ and the relative bearing vector as $ g_{ij} = \frac{z_{ij}}{d_{ij}} \in \RN^{2} $, all relative to a global coordinate frame $ \Sigma^{\text{g}} $. 
It follows $ z_{ji} = - z_{ij} $, $ d_{ji} = d_{ij} $ and $ g_{ji} = - g_{ij} $.

\subsubsection{Distance-based formation control}
In \textit{distance-based} formation control, a \textit{desired formation} is characterized by a set of \textit{inter-robot distance constraints}.
Assume the desired distance between a robot pair $ \BR{i, \, j} $ of the formation is $ d_{ij}^{\star} $ and let $ d_{ij}\BR{t} $ be the current distance at time $ t $.
Let us define the \textit{distance error} signal as $ e_{ij\text{d}}\BR{t} = d_{ij}^{2}\BR{t} - {\BR{d_{ij}^{\star}}}^{2} $.
A distance-based potential function used for deriving the gradient-based control law for the robot pair $ \BR{i, \, j} $ takes the form $ V_{ij{\text{d}}}\BR{e_{ij\text{d}}} = \frac{1}{4} e_{ij{\text{d}}}^{2} $.
It has a minimum at the desired edge distance $ d_{ij}^{\star} $; in other words, $ V_{ij\text{d}}\BR{e_{ij\text{d}}} \geq 0 $ and $ V_{ij\text{d}}\BR{e_{ij\text{d}}} = 0 \iff e_{ij{\text{d}}} = 0 \iff d_{ij} = d_{ij}^{\star} $. 
In this case, the corresponding gradient-based control law for maintaining a desired inter-robot distance for the robot pair $ \BR{i, \, j} $ is
\[ 
u_{ij\text{d}} = -{\BR{\PARDIFF{V_{ij\text{d}}}{p_{i}}}}^{\top} = e_{ij\text{d}} z_{ij},
\]
where $ z_{ij} $ is the measurement that \texttt{Ri} obtains from its neighbor $ j \in \mathcal{N}_{i} $. 
Thus the distanced-based formation control law for robot \texttt{Ri} in \eqref{eq:nR-Robot-Dynamics} is given by 
\begin{equation} \label{eq:dist-gradient-control}
    u_{i\text{d}} = \sum_{j\in \mathcal{N}_{i}} e_{ij\text{d}} z_{ij}.
\end{equation}
It is well-studied in the literature that the above control law guarantees the local exponential stability of the desired formation shape when the desired shape is infinitesimally rigid. 
We refer interested readers to \cite{Sun2016} for the exposition of standard distance-based formation control and its local exponential stability property.  

\subsubsection{Bearing-only formation control}
In \textit{bearing-only} formation control, the desired formation is characterized by a set of \textit{inter-robot bearing} constraints. 
Consider the $ i $-th robot (with label \texttt{Ri}) in this setup. 
Robot \texttt{Ri} is able to obtain the bearing measurement $ g_{ij}\BR{t} $ to its neighbors $ j \in \mathcal{N}_{i} $ and its goal is to achieve desired bearings $ g_{ij}^{\star} $s with all neighbors $ j \in \mathcal{N}_{i} $.  
In this case, the \textit{bearing error} signal for a robot pair $ \BR{i, \, j} $ can be defined by $ e_{ij\text{b}}\BR{t} = g_{ij}\BR{t} - g_{ij}^{\star} $. 
As before, the corresponding potential function that can be used to design the gradient-based control law is $ V_{ij\text{b}}\BR{e_{ij\text{b}}} = d_{ij} \norm{e_{ij\text{b}}}^{2} $. 
Note that $ V_{ij\text{b}}\BR{e_{ij\text{b}}} \geq 0 $ and it is only zero when $ d_{ij} = 0 $ or $ e_{ij\text{b}} = \mathbb{0}_{2} \iff g_{ij} = g_{ij}^{\star} $. 
(In forthcoming analysis, we will show that $ d_{ij} = 0 $, where robots \texttt{Ri} and \texttt{Rj} are at the same position, is not a viable option.)
It can be verified that 
\[
u_{ij\text{b}} = -{\BR{\PARDIFF{V_{ij\text{b}}}{p_{i}}}}^{\top} = e_{ij\text{b}} 
\]
is the gradient-based control law derived from $ V_{ij\text{b}}\BR{e_{ij\text{b}}} $ for the robot pair $ \BR{i, \, j} $. 
The bearing-only formation control law for robot \texttt{Ri} in \eqref{eq:nR-Robot-Dynamics} is then given by
\begin{equation} \label{eq:bearing-gradient-control}
    u_{i\text{b}} = \sum_{j\in\mathcal{N}_i}e_{ij\text{b}}.
\end{equation}
In \cite{Zhao2018}, it has been shown that the above control law ensures the global asymptotic stability of the desired formation shape provided the formation shape is infinitesimally bearing rigid. 

\subsection{Cubic equations} \label{subsec:Cubic-Equations}

Consider the \textit{reduced}\footnote{in some texts, the term `depressed' is used.} cubic equation
\begin{equation} \label{eq:Prel-Reduced-Cubic}
    y^{3} + cy + d = 0
    ,
\end{equation}
where $ c, \, d \in \RN $.
The discriminant of \eqref{eq:Prel-Reduced-Cubic} is $ \Delta = -4 c^{3} - 27 d^{2} $.
Using $ \Delta $, we determine the following properties regarding the roots:
\begin{itemize}
    \item 
    $ \Delta > 0 $; we have three distinct real roots;
    
    \item 
    $ \Delta = 0 $; we have at least two equal real roots;
    
    \item 
    $ \Delta < 0 $; we have a single real root and two complex roots forming a conjugate pair.
\end{itemize}
The roots of $ \eqref{eq:Prel-Reduced-Cubic} $ are \cite{Dickson1922}
\begin{equation} \label{eq:Prel-Reduced-Cubic-Roots}
    \begin{aligned}
        y_{1} = A + B
        , \quad 
        y_{2} = \omega A + \omega^{2} B
        , \quad 
        y_{3} = \omega^{2} A + \omega B
        , 
    \end{aligned}
\end{equation}
where
\begin{equation} \label{eq:Prel-Reduced-Cubic-Roots-Parameters}
    \begin{aligned}
        &
        A = \sqrt[3]{-\frac{d}{2} + \sqrt{R}}
        ,
        & & 
        B = \sqrt[3]{-\frac{d}{2} - \sqrt{R}}
        ,
        \\
        & 
        AB = -\frac{c}{3}
        ,
        & & 
        R = {\BR{\frac{c}{3}}}^{3} + {\BR{\frac{d}{2}}}^{2} 
        = \frac{-1}{108} \Delta
        ,
        \\
        & 
        \omega = \frac{-1 + \sqrt{3} \, \text{i}}{2}
        ,
        & & 
        \omega^{2} = \frac{-1 - \sqrt{3} \, \text{i}}{2}
        .
    \end{aligned}
\end{equation}
Note that $ \omega $ and $ \omega^{2} $ form a complex conjugate pair, i.e, $ \widebar{\omega} = \omega^{2} $.
In polar form, we obtain $ \omega = 1 \angle 120^{\degree} $ and $ \omega^{2} = 1 \angle \BR{-120^{\degree}} $.

\vspace{0.5\baselineskip}
We focus on the case when the coefficients of \eqref{eq:Prel-Reduced-Cubic} take values $ c < 0 $ and $ d \gtreqqless 0 $. 
Applying Descartes' rule of signs \cite{Dickson1922}, we obtain that when the coefficient $ d \leq 0 $, the reduced cubic equation \eqref{eq:Prel-Reduced-Cubic} always has a positive real root while for $ d > 0 $, it always contains a negative real root. 
The remaining roots for the case when $ d $ is positive depends on the discriminant $ \Delta $; when $ \Delta \geq 0 $, we have two positive real roots, and otherwise, we have zero positive real roots and instead two complex roots. 
The following lemma provides the characterization of the positive real roots when $ d > 0 $.

\vspace{0.5\baselineskip}
\begin{lemma} \label{lem:Prel-Reduced-Cubic-Roots-Positive}
    Consider the reduced cubic equation \eqref{eq:Prel-Reduced-Cubic} with coefficients $ c < 0 $ and $ d > 0 $. 
    Assume the discriminant is $ \Delta \geq 0 $. 
    Then two positive real roots exist with values
    \begin{equation} \label{eq:Prel-Reduced-Cubic-Roots-Positive}
        \begin{aligned}
            y_{\text{p}_{1}} 
                & = 2 \sqrt[3]{r_{v}} \cos \BR{\frac{1}{3} \varphi_{v}}
                & & \in \left[1, \, \sqrt{3}\right) \sqrt[3]{r_{v}}
                ,
                \\
            y_{\text{p}_{2}} 
                & = 2 \sqrt[3]{r_{v}} \cos \BR{\frac{1}{3} \varphi_{v} - 120^{\degree}}
                & & \in \left(0, \, 1\right] \sqrt[3]{r_{v}}
            ,
        \end{aligned}
    \end{equation} 
    where $ r_{v} = \sqrt{- \BR{\frac{c}{3}}^{3}} $ and $ \varphi_{v} = \tan^{-1} \BR{\frac{-2}{d}\sqrt{-R}} \in \left(90^{\degree}, \, 180^{\degree} \right] $.
    When $ \Delta = 0 $, the two positive real roots are equal and have value $ y_{\text{p}_{1}} = y_{\text{p}_{2}} = \sqrt[3]{r_{v}} = \sqrt[3]{\frac{d}{2}} $.
\end{lemma}

\vspace{0.5\baselineskip}
The proof of Lemma \ref{lem:Prel-Reduced-Cubic-Roots-Positive} is found in Appendix \ref{subsec:App-Lemma-Proof-1}.

\subsection{Problem formulation} \label{subsec:Problem-Setup}
As discussed in the Introduction, we study in this paper the setup in which the robots possess heterogeneous sensing, and each robot, depending on its own local information, maintains the prescribed distance or bearing with its neighbors using the aforementioned distance-based or bearing-only formation control law.
Thus, in the current setup, each robot fulfills either a \textit{distance task} or a \textit{bearing task}. 
As before, consider a pair of robots with labels \texttt{Ri} and \texttt{Rj}. 
In case \texttt{Ri} is assigned a \textit{distance task}, its goal is to maintain a desired distance $ d_{ij}^{\star} \in \RN_{> 0} $ with \texttt{Rj}. 
The robot \texttt{Ri} possesses an \textit{independent local coordinate frame} $ \Sigma^{\text{i}} $ which is \textit{not necessarily aligned} with that of \texttt{Rj} or the \textit{global coordinate frame} $ \Sigma^{\text{g}} $.
Within its local coordinate frame $ \Sigma^{\text{i}} $, \texttt{Ri} is able to measure the relative position vector $ z_{ij} \in \RN^{2} $ relative to \texttt{Rj}.
On the other hand, when \texttt{Rj} is assigned a \textit{bearing task}, its goal is to maintain a desired bearing $ g_{ji}^{\star} \in \RN^{2} $ with \texttt{Ri}. The robot 
\texttt{Rj} is able to obtain the relative bearing measurement $ g_{ji} $ of \texttt{Ri} in its \textit{local coordinate frame} $ \Sigma^{\text{j}} $ which is \textit{aligned} with $ \Sigma^{\text{g}} $.
Since robot \texttt{Ri} is assigned the distance task, we term it a \textit{distance robot}. 
Correspondingly, \texttt{Rj} is a \textit{bearing robot}. 
For the interconnection topology, we assume each robot has only neighbors of the opposing category, i.e., a distance robot can only have edges with bearing robot(s) and vice versa. 
As a result, the team of $ n $ robots can be partitioned into two sets, namely the set of distance robots $ \mathcal{D} $ and the set of bearing robots $ \mathcal{B} $ with $ \mathcal{D} \neq \emptyset $ and $ \mathcal{B} \neq \emptyset $.
The edge set is given by $ \mathcal{E} \subseteq \BR{\mathcal{D} \times \mathcal{B}} \cup \BR{\mathcal{B} \times \mathcal{D}} $; the underlying graph structure is that of a bipartite digraph. 

\begin{figure}[!tb]
    \centering
    \includegraphics[width=0.95\linewidth]{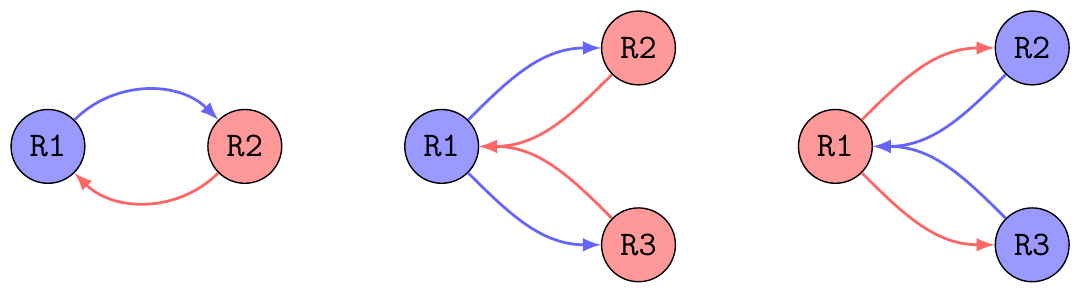}
    \caption{Setups for the two and three robot case; \protect\tikz{\protect\draw[fill = blue!40!white] (0, 0) circle [radius = 4pt];}  represents a distance robot and \protect\tikz{\protect\draw[fill = red!40!white] (0, 0) circle [radius = 4pt];} represents a bearing robot.
    Correspondingly, blue arrows are sensing carried out by the distance robots \protect\tikz{\protect\draw[fill = blue!40!white] (0, 0) circle [radius = 4pt];} while red arrows represents the edges from the bearing robots \protect\tikz{\protect\draw[fill = red!40!white] (0, 0) circle [radius = 4pt];}.
    From left to right, we have the (\textbf{1D1B}), (\textbf{1D2B}), and (\textbf{1B2D}) setup.
    }
    \label{fig:Robot-Topology}
\end{figure}

In the current work, we focus on the cases in which the team of $ n \in \CBR{2, \, 3} $ robots has a \textit{complete bipartite digraph} topology, i.e., the edge set is $ \mathcal{E} = \BR{\mathcal{D} \times \mathcal{B}} \cup \BR{\mathcal{B} \times \mathcal{D}} $. 
For the two robot case, we have only one feasible setup, namely the setup consisting of \textit{one distance and one bearing} (\textbf{1D1B}) robot, while for the three robot case we have two feasible robot setups, namely the \textit{one distance and two bearing} (\textbf{1D2B}) or the \textit{one bearing and two distance} (\textbf{1B2D}) setup; see Fig. \ref{fig:Robot-Topology} for an illustration of these setups. 
Based on these setups, we are interested in studying the stability of the formation when the distance-based formation control law in \eqref{eq:dist-gradient-control} for the distance robot(s) and the bearing-only formation control law in \eqref{eq:bearing-gradient-control} for the bearing robot(s) are used. 
In this case, we do not modify the standard gradient-based control law for the different tasks. 
Consequently, we analyze whether i). the equilibrium set contains undesired shape and/or group motion; ii). the desired shape is (exponentially) stable; and iii). the undesired shape and/or group motion (if any) is attractive. 
The first and last questions are motivated by the robustness issues of the distance- and displacement-based controllers as studied in \cite{Belabbas2012, Mou2016, Sun2017, deMarina2021} where a disagreement between neighboring robots about desired values or measurements can lead to an undesired group motion and deformation of the formation shape. 
Since we are considering heterogeneous sensing mechanisms with corresponding heterogeneous potential functions, it is of interest whether such undesired behaviour can co-exist. 
Such knowledge on the effect of heterogeneity in the control law can potentially be useful to design simultaneous formation and motion controller as pursued recently in \cite{Marina2016}.

\section{THE (\textbf{1D1B}) ROBOT SETUP} \label{sec:The-1D1B-Setup}
In this section, we focus on the case of two robots in the (\textbf{1D1B}) setup as depicted in Fig. \ref{fig:Robot-Topology}. 
The analysis of this seemingly simple setup serves as a prelude for the setups with three robots. 
Without loss of generality, robot \texttt{R1} takes the role of the distance robot and \texttt{R2} is the bearing robot.
Considering gradient-based control laws for the robot-specific tasks, the closed-loop dynamics is given by 
\begin{equation} \label{eq:2R-1D1B-Closed-Loop}
    \begin{aligned}
        \dot{p}_{1} 
            & = K_{\text{d}} e_{12\text{d}} z_{12}
            \\
        \dot{p}_{2}
        & = K_{\text{b}} e_{21\text{b}}
    \end{aligned}
    \iff 
    \begin{aligned}
        \dot{p}
        =
        \begin{bmatrix}
            K_{\text{d}} z_{12} & \mathbb{0}_{2 \times 2}
            \\
            \mathbb{0}_{2 \times 1} & - K_{\text{b}} I_{2}
        \end{bmatrix}
        \dot{e}
        ,
    \end{aligned}    
\end{equation}
where $ K_{\text{d}} > 0 $ and $ K_{\text{b}} > 0 $ are control gains for robots \texttt{R1} and \texttt{R2}, respectively, the bearing error satisfies $ e_{21\text{b}} = - e_{12\text{b}} $ in $ \Sigma^{\text{g}} $, and the error vector is $ e = {\begin{bmatrix} e_{12\text{d}} & e_{12\text{b}}^{\top} \end{bmatrix}}^{\top} \in \RN^{3} $.
It is of interest to note that when physical dimension is taken into account with $ \left[\text{L}\right] $ as the unit of length and $ \left[\text{T}\right] $ the unit of time, the control gain $ K_{\text{d}} $ has dimension $ \left[ \text{L}\right]^{-2}\left[ \text{T}\right]^{-1} $ while $ K_{\text{b}} $ is expressed in $ \left[ \text{L} \right]\left[\text{T}\right]^{-1} $. 
This observation implies that the ratio of both gains has an effect on the time-scale of both systems' dynamics.

In the remainder of this section, we provide the analysis of the closed-loop formation system \eqref{eq:2R-1D1B-Closed-Loop}, hereby focusing on the three questions raised in Section \ref{subsec:Problem-Setup}.
First, we have the following result on the equilibrium configurations.

\vspace{0.5\baselineskip}
\begin{proposition}[\emph{(\textbf{1D1B})} Equilibrium Configurations] \label{prop:2R-1D1B-Equilibrium_Configurations}
    The equilibrium configurations for the closed-loop formation system \eqref{eq:2R-1D1B-Closed-Loop} belong to the set 
    \begin{equation} \label{eq:2R-1D1B-Equilibrium-Set}
        \mathcal{S}_{p} = \CBR{p \in \RN^{4} \, \rvert \, e = \mathbb{0}_{3}}
        .
    \end{equation}
\end{proposition}

\vspace{0.5\baselineskip}
\begin{IEEEproof}
    Solving for $ \dot{p}_{1} = \mathbb{0}_{2} $ and $ \dot{p}_{2} = \mathbb{0}_{2} $, we obtain $ K_{\text{d}} e_{12\text{d}} z_{12} = \mathbb{0}_{2} \iff e_{12\text{d}} = 0 \vee z_{12} = \mathbb{0}_{2} $ and $ - K_{\text{b}} e_{12\text{b}} = \mathbb{0}_{2} \iff g_{12} = g_{12}^{\star} $, respectively.
    With $ g_{12} = g_{12}^{\star} $ implying $ d_{12} \neq 0 $, it follows the option $ z_{12} = \mathbb{0}_{2} $ is not feasible.
    Hence, $ \dot{p} = \mathbb{0}_{4} \iff e = \mathbb{0}_{3} $.
    This completes the proof.
\end{IEEEproof}

\vspace{0.5\baselineskip}
Following $ e = \mathbb{0}_{3} $, the inter-robot relative position $ z_{12} $ equals the desired relative position $ z_{12}^{\star} = d_{12}^{\star} g_{12}^{\star} $ when both the robots attain their individual task. 
Furthermore, note the set $ \mathcal{S}_{p} $ is invariant under translations in the plane; therefore, such a set is non-compact. 

We proceed the analysis by determining the stability of the equilibrium configurations \eqref{eq:2R-1D1B-Equilibrium-Set}. 
To this end, we take as Lyapunov function candidate $ V\BR{e} $ of the form
\begin{equation} \label{eq:2R-1D1B-Lyapunov-Function}
    \begin{aligned}
        V\BR{e} 
            & = V_{{12\text{d}}} \BR{e_{12\text{d}}} + V_{{21\text{b}}} \BR{e_{21\text{b}}}
            \\
            & = \frac{1}{4} K_{\text{d}} e_{12\text{d}}^{2} + \frac{1}{2} K_{\text{b}} d_{21} \norm{e_{21\text{b}}}^{2}
            .
    \end{aligned}
\end{equation}
Observe that $ V\BR{e} $ is the sum of the task-specific potential functions. 
Since $ V_{{12\text{d}}} \BR{e_{12\text{d}}} \geq 0 $ and $ V_{{21\text{b}}} \BR{e_{21\text{b}}} \geq 0 $, it follows that $ V\BR{e} \geq 0 $.
Moreover, $ V\BR{e} = 0 \iff V_{{12\text{d}}} \BR{e_{12\text{d}}} = 0 \, \wedge \, V_{{21\text{b}}} \BR{e_{21\text{b}}} = 0 $.
When considered separately, we know $ V_{{12\text{d}}} \BR{e_{12\text{d}}} = 0 \iff d_{12} = d_{12}^{\star} $ and $ V_{{21\text{b}}} \BR{e_{21\text{b}}} = 0 \iff d_{21} = 0 \vee e_{21\text{b}} = \mathbb{0}_{2} $. 
Combining both potential functions, we conclude that $ d_{21} = 0 $ is not a feasible option since then $ V_{12\text{d}} \BR{e_{12\text{d}}} > 0 $. 
Therefore, the minimum value of $ V\BR{e}$ is attained in $ \mathcal{S}_{p} $, i.e., $  V\BR{e} = 0 \iff \widehat{p} \in \mathcal{S}_{p} $.
The derivative of \eqref{eq:2R-1D1B-Lyapunov-Function} evaluates to
\begin{equation} \label{eq:2R-1D1B-Lyapunov-Function-Derivative}
    \begin{aligned}
        \dot{V} \BR{e}
            & = \DIFF{}{t} \big( V_{{12\text{d}}}\BR{e_{12\text{d}}} + V_{{21\text{b}}}\BR{e_{21\text{b}}} \big)
            \\
            & = 
            \begin{bmatrix}
                \PARDIFF{}{p_{1}} V_{{12\text{d}}} & \PARDIFF{}{p_{2}} V_{{12\text{d}}} 
            \end{bmatrix}
            \begin{bmatrix}
                \dot{p}_{1}
                \\
                \dot{p}_{2}
            \end{bmatrix}
            \\
            & 
            \qquad +
            \begin{bmatrix}
                \PARDIFF{}{p_{1}} V_{{21\text{b}}} & \PARDIFF{}{p_{2}} V_{{21\text{b}}} 
            \end{bmatrix}
            \begin{bmatrix}
                \dot{p}_{1}
                \\
                \dot{p}_{2}
            \end{bmatrix}
            \\
            & = 
            \begin{bmatrix}
                - \dot{p}_{1}^{\top} & \dot{p}_{1}^{\top}
            \end{bmatrix}
            \begin{bmatrix}
                \dot{p}_{1}
                \\
                \dot{p}_{2}
            \end{bmatrix}
            +
            \begin{bmatrix}
                \dot{p}_{2}^{\top} & - \dot{p}_{2}^{\top}
            \end{bmatrix}
            \begin{bmatrix}
                \dot{p}_{1}
                \\
                \dot{p}_{2}
            \end{bmatrix}
            \\
            & = - \norm{\dot{z}_{12}}^{2}
            \leq 0
            ,
    \end{aligned}
\end{equation}
where we use $ \dot{p}_{i} = - {\BR{\PARDIFF{}{p_{i}} V_{{ij\bullet}}}}^{\top} \iff - p_{i}^{\top} = \PARDIFF{}{p_{i}} V_{{ij\bullet}} $ and $ \PARDIFF{}{p_{j}} V_{{ij\bullet}} = - \PARDIFF{}{p_{i}} V_{{ij\bullet}} $ for $ i, \, j \in \CBR{1, \, 2} $ and $ \bullet \in \CBR{\text{b}, \, \text{d}} $.
From \eqref{eq:2R-1D1B-Lyapunov-Function-Derivative}, we have $ \dot{V}\BR{e} $ is negative semi-definite and $ \dot{V} \BR{e} = 0 \iff \dot{p}_{1} = \dot{p}_{2} $.
The following are invariant sets corresponding to $ \dot{V}\BR{e} = 0 $:
\begin{enumerate}
    \item 
    $ \dot{p}_{1} = \dot{p}_{2} = \mathbb{0}_{2} $; these are the previously obtained equilibrium configurations $ \widehat{p} \in \mathcal{S}_{p} $;
    
    \item 
    $ \dot{p}_{1} = \dot{p}_{2} = w \neq \mathbb{0}_{2} $; in this scenario, the robots move with a (yet to be determined) constant velocity $ w \in \RN^{2} $.
\end{enumerate}

Let the set of configurations $ p \in \RN^{4} $ yielding robots to move with the (yet to be determined) constant velocity $ w $ be given as  
\begin{equation} \label{eq:2R-1D1B-Moving-Set}
    \mathcal{T}_{p} = \CBR{p \in \RN^{4} \, \rvert \, \dot{p}_{1} = \dot{p}_{2} = w, \, w \neq \mathbb{0}_{2}}
    .
\end{equation}
Note the set $ \mathcal{T}_{p} $ is also non-compact, since if $ \widehat{p} \in \mathcal{T}_{p} $, then $ \widehat{p} + c \BR{\mathbb{1}_{2} \otimes w} \in \mathcal{T}_{p} $ with $ c \in \RN $.
Since both the equilibrium set $ \mathcal{S}_{p} $ and the \textit{moving} set $ \mathcal{T}_{p} $ are non-compact and the expression for the Lyapunov derivative \eqref{eq:2R-1D1B-Lyapunov-Function-Derivative} is expressed in the dynamics of the \textit{link} $ z_{12} $, we continue the analysis by exploring the \textit{link dynamics} $ \dot{z}_{12} $, obtained as
\begin{equation} \label{eq:2R-1D1B-Link-Dynamics}
    \begin{aligned}
        \dot{z}_{12} = - \BR{K_{\text{b}} e_{12\text{b}} + K_{\text{d}} e_{12\text{d}} z_{12}}
        ,
    \end{aligned}
\end{equation}
instead of the robot dynamics $ \dot{p} $. 
Mapping the sets of interest $ \mathcal{S}_{p} $ and $ \mathcal{T}_{p} $ to the \textit{link space} yields 
$ \mathcal{S}_{z} = \CBR{z_{12} \in \RN^{2} \, \rvert \, z_{12} = z_{12}^{\star}} $ and $ \mathcal{T}_{z} = \CBR{z_{12} \in \RN^{2} \, \rvert \, z_{12} = z_{12_{\text{M}}}} $, where $ z_{12_{\text{M}}} = d_{12_{\text{M}}} g_{12_{\text{M}}} $ is the inter-robot relative position vector as they move with the (yet unknown) velocity $ w $ in the plane.
The set $ \mathcal{S}_{z} $ contains only a single point and is therefore compact. 

\subsection{Characterization of the moving set $ \mathcal{T}_{z} $}
In this part, we make the effort to characterize the moving set $ \mathcal{T}_{z} $ (and implicitly $ \mathcal{T}_{p} $).
To this end, we provide the following proposition:

\vspace{0.5\baselineskip}
\begin{proposition}[\emph{(\textbf{1D1B})} Moving Configurations]  \label{prop:2R-1D1B-Moving_Configurations}
    The closed-loop formation system \eqref{eq:2R-1D1B-Closed-Loop} moves with a constant velocity $ w = 2 K_{\text{b}} g_{12}^{\star} $ when the error vector $ e $ is of the form
    \begin{equation}
        \begin{bmatrix}
            e_{12\text{d}}
            \\
            e_{12\text{b}}
        \end{bmatrix}
        =
        -
        \begin{bmatrix}
            \frac{2}{d_{12}} R_{\text{bd}}
            \\
            2 g_{12}^{\star}
        \end{bmatrix}
        .
    \end{equation}
\end{proposition}

\vspace{0.5\baselineskip}
\begin{IEEEproof}
    Solving for $ \dot{z}_{12} = \mathbb{0}_{2} $ results in the expression $ \BR{K_\text{d} e_{12\text{d}} d_{12} + K_\text{b}} g_{12} = K_\text{b} g_{12}^{\star} $.
    We distinguish two cases.
    The case $ g_{12} = g_{12}^{\star} $ and $ K_\text{d} e_{12\text{d}} d_{12} + K_\text{b} = K_\text{b} \iff K_\text{d} e_{12\text{d}} d_{12} = 0 $ corresponds to the equilibrium configurations $ \widehat{p} \in \mathcal{S}_{p} $.
    On the other hand, we have $ g_{12} = - g_{12}^{\star} $ and $ K_\text{d} e_{12\text{d}} d_{12} + K_\text{b} = - K_\text{b} \iff e_{12\text{d}} d_{12} + 2 R_{\text{bd}} = 0 $, where $ R_{\text{bd}} = \frac{K_{\text{b}}}{K_{\text{d}}} $ is the ratio of the control gains.
    Substituting in the robot dynamics \eqref{eq:2R-1D1B-Closed-Loop} yields $ \dot{p}_{1} = \dot{p}_{2} = 2 K_{\text{b}} g_{12}^{\star} $, i.e., robots are moving with a common velocity $ w = 2 K_{\text{b}} g_{12}^{\star} $. 
    Therefore, moving formations occur at the relative orientation $ g_{12} = - g_{12}^{\star} $.
    By definition, we obtain $ e_{12\text{b}} = -2 g_{12}^{\star} $ and the distance error $ e_{12\text{d}} d_{12} + 2 R_{\text{bd}} = 0 \iff e_{12\text{d}} = -\frac{2}{d_{12}} R_{\text{bd}} $.
    This completes the proof.
\end{IEEEproof}

\vspace{0.5\baselineskip}
Proposition \ref{prop:2R-1D1B-Moving_Configurations} provides a characterization of the moving set $ \mathcal{T}_{z} $ through the error vector $ e $. 
We also obtain that the inter-robot bearing for moving formations to occur is $ g_{12_{\text{M}}} = - g_{12}^{\star} $.
For a complete characterization of $ \mathcal{T}_{z} $ in terms of the vector $ z_{12_{\text{M}}} $, it remains to obtain the distance $ d_{12_{\text{M}}} $ by solving the expression $ e_{12\text{d}} d_{12} + 2 R_{\text{bd}} = 0 $.
Expanding it, we obtain the following cubic equation in $ d_{12} $:
\begin{equation} \label{eq:2R-1D1B-Cubic-Equation-d12}
    \begin{aligned}
        d_{12}^{3} - {\BR{d_{12}^{\star}}}^{2} d_{12} + 2 R_{\text{bd}}
            & = 0
            .
    \end{aligned}
\end{equation}
Compared with \eqref{eq:Prel-Reduced-Cubic}, the coefficients are $ c = - {\BR{d_{12}^{\star}}}^{2} < 0 $ and $ d = 2 R_{\text{bd}} > 0 $.
We infer the solution set to \eqref{eq:2R-1D1B-Cubic-Equation-d12} contains positive real roots given by Lemma \ref{lem:Prel-Reduced-Cubic-Roots-Positive} when the corresponding discriminant is non-negative.
This is equivalent to the constraint $ d_{12}^{\star} \geq \widehat{d} $ with $ \widehat{d} = \sqrt{3} R_{\text{bd}}^{\frac{1}{3}} $.

\vspace{0.5\baselineskip}
\begin{remark} \label{rem:2R-1D1B-Global-Stability}
    When the desired distance $ d_{12}^{\star} < \widehat{d} $, the moving set $ \mathcal{T}_{z} = \emptyset $ since there does not exist a feasible distance between robots such that they move with the common velocity $ w $. 
    This implies $ \dot{V}\BR{e} = 0 \iff \widehat{p} \in \mathcal{S}_{p} $. 
    With $ \dot{V}\BR{e} < 0 $ when $ p \not \in \mathcal{S}_{p} $, we conclude that for all initial configurations $ p\BR{0} \in \RN^{4} $ satisfying $ p_{1}\BR{0} \neq p_{2}\BR{0} $, we have global asymptotic convergence to the desired equilibrium set $ \mathcal{S}_{p} $.
\end{remark}

\vspace{0.5\baselineskip}
\begin{remark} \label{rem:2R-1D1B-Delay-Tp}
    The threshold value $ \widehat{d} $ is proportional to the gain ratio $ R_{\text{bd}}^{\frac{1}{3}} $; increasing $ R_{\text{bd}} $ leads to a larger value for $ \widehat{d} $.
    Therefore, increasing the value of $ R_{\text{bd}} $ ``delays" the occurrence of the moving set $ \mathcal{T}_{p} $ since there exists a larger range of values for $ d_{12}^{\star} $ satisfying $ d_{12}^{\star} < \widehat{d} $.
\end{remark}

\vspace{0.5\baselineskip}
Assume the desired distance satisfies $ d_{12}^{\star} \geq \widehat{d} $.
Lemma \ref{lem:Prel-Reduced-Cubic-Roots-Positive} provides us the positive roots and thus the feasible distances $ d_{12} $ satisfying \eqref{eq:2R-1D1B-Cubic-Equation-d12}. 
For the specific values $ c = - {\BR{d_{12}^{\star}}}^{2} $ and $ d = 2 R_{\text{bd}} $, we obtain $ r_{v} = \sqrt{\frac{{\BR{d_{12}^{\star}}}^{6}}{27}} $ and $ \varphi_{v} = \tan^{-1} \BR{- R_{\text{bd}}^{-1} \sqrt{-R}} $.
Substituting in \eqref{eq:Prel-Reduced-Cubic-Roots-Positive} yields
\begin{equation} \label{eq:2R-1D1B-Cubic-Equation-Roots-Positive}
    \begin{aligned}
        y_{\text{p}_{1}} 
            & = \frac{2}{3} \sqrt{3} d_{12}^{\star} \cos \BR{\frac{1}{3} \varphi_{v}}
            ,
            \\
        y_{\text{p}_{2}} 
            & = \frac{2}{3} \sqrt{3} d_{12}^{\star} \cos \BR{\frac{1}{3} \varphi_{v} - 120^{\degree}}
            .
    \end{aligned}
\end{equation}
When $ d_{12}^{\star} = \widehat{d} $, the positive root (with multiplicity $ 2 $) corresponding to the cubic equation \eqref{eq:2R-1D1B-Cubic-Equation-d12} is $ d_{12} = \frac{1}{3} \sqrt{3} \widehat{d} = R_{\text{bd}}^{\frac{1}{3}} $.
The characterization of the moving set $ \mathcal{T}_{z} $ for $ d_{12}^{\star} \geq \widehat{d} $ is then
\begin{equation} \label{eq:2R-1D1B-Tz-Delta>0}
    \begin{aligned}
        \mathcal{T}_{z}
            =
            \CBR{z_{12} \in \RN^{2} \, \rvert \, z_{12} = - y g_{12}^{\star}, \, y \text{ satisfies } \eqref{eq:2R-1D1B-Cubic-Equation-Roots-Positive}}
            .
    \end{aligned}
\end{equation}

\subsection{Local stability of the moving set $ \mathcal{T}_{z} $}
After characterizing the set $ \mathcal{T}_{z} $ for $ d_{12}^{\star} \geq \widehat{d} $, we continue with determining the local stability of $ \mathcal{T}_{z} $.
First, we obtain the Jacobian of the right hand side (RHS) of the link dynamics \eqref{eq:2R-1D1B-Link-Dynamics} as the matrix
\begin{equation} \label{eq:2R-1D1B-Jacobian}
    \begin{aligned}
        A 
            & = - \BR{K_{\text{b}} A_{12\text{b}} + K_{\text{d}} A_{12\text{d}}}
            ,
    \end{aligned}
\end{equation}
where $ A_{12\text{b}} = \frac{1}{d_{12}} \BR{I - g_{12} g_{12}^{\top}} $ and $ A_{12\text{d}} = e_{12\text{d}}I + 2 z_{12} z_{12}^{\top} $.

\vspace{0.5\baselineskip}
\begin{lemma} \label{lem:2R-1D1B-Jacobian-Tz}
    The closed-loop formation system \eqref{eq:2R-1D1B-Closed-Loop} at any feasible point $ \widehat{z}_{12} $ in the moving set $ \mathcal{T}_{z} $ is unstable.  
\end{lemma}

\vspace{0.5\baselineskip}
\begin{IEEEproof}
    From Proposition \ref{prop:2R-1D1B-Moving_Configurations}, we know that all feasible points $ \widehat{z}_{12} \in \mathcal{T}_{z} $ satisfies $ e_{12\text{d}} = -2 \frac{1}{d_{12}} R_{\text{bd}} < 0 $ and $ g_{12} = -g_{12}^{\star} $. 
    Let the inter-robot bearing be given as $ g_{12} = {\begin{bmatrix} a & b \end{bmatrix}}^{\top} $ with the norm constraint $ \norm{g_{12}} = \sqrt{a^{2} + b^{2}} = 1 $; we obtain $ g_{12}g_{12}^{\top} = \left[\begin{smallmatrix} a^{2} & ab \\ ab & b^{2} \end{smallmatrix} \right] = g_{12}^{\star} g_{12}^{\star \, \top} $.
    Substituting in \eqref{eq:2R-1D1B-Jacobian} yields
    \begin{equation} \label{eq:2R-1D1B-Jacobian-Tz}
        \begin{aligned}
            A
                & = 
                \begin{bmatrix}
                    x + \BR{x - y} a^{2} & \BR{x - y} ab
                    \\
                    \BR{x - y} ab &  x + \BR{x - y} b^{2}
                \end{bmatrix}
                ,
        \end{aligned}
    \end{equation}
    where we define $ x = K_{\text{b}} \frac{1}{d_{12}} $ and $ y = 2 K_{\text{d}} d_{12}^{2} $, both being positive.
    For a $ 2 \times 2 $ matrix $ A $, the characteristic polynomial $ \chi\BR{\lambda} $ can be given by 
    \begin{equation} \label{eq:2R-1D1B-Jacobian-Tz-Char-Pol}
        \chi\BR{\lambda} = \lambda^{2} - \TRACE{A} \lambda + \DET{A},
    \end{equation}
    where $ \TRACE{A} $ is the trace and $ \DET{A} $ is the determinant of matrix $ A $.
    With $ A $ in \eqref{eq:2R-1D1B-Jacobian-Tz}, we obtain $ \TRACE{A} = 3 x - y $ and $ \DET{A} = x\BR{2x - y} $. 
    It follows the roots of $ \chi \BR{\lambda} $ are $ \lambda_{1} = x $ and $ \lambda_{2} = 2x - y $. 
    Since $ \lambda_{1} = x > 0 $, we conclude matrix $ A $ has at least one positive eigenvalue for all feasible points $ \widehat{z}_{12} \in \mathcal{T}_{z} $.
    Therefore, the closed-loop formation system \eqref{eq:2R-1D1B-Closed-Loop} is unstable at these points.
    This completes the proof. 
\end{IEEEproof}

\vspace{0.5\baselineskip}
The following theorem states the main result for the (\textbf{1D1B}) setup when the desired distance satisfies $ d_{12}^{\star} \geq \widehat{d} $.

\vspace{0.5\baselineskip}
\begin{theorem}[Almost global convergence] \label{thm:2R-1D1B-Almost-Global-Convergence}
    Consider two robots \texttt{R1} and \texttt{R2} possessing heterogeneous sensing mechanisms. 
    Let the closed-loop dynamics be given by \eqref{eq:2R-1D1B-Closed-Loop} and the desired inter-robot distance satisfies $ d_{12}^{\star} \geq \widehat{d} $. 
    If $ p_{1}\BR{0} \neq p_{2}\BR{0} $ and $ p\BR{0} \not \in \mathcal{T}_{p} $, then the trajectories of the robots asymptotically converge to a point $ \widehat{p} \in \mathcal{S}_{p} $.
\end{theorem}

\vspace{0.5\baselineskip}
\begin{IEEEproof}
    In \eqref{eq:2R-1D1B-Lyapunov-Function-Derivative}, we obtain the Lyapunov derivative $ \dot{V}\BR{e} \leq 0 $. In fact, $ \dot{V}\BR{e} < 0 $ when $ p\BR{t} \not\in \mathcal{S}_{p} \cup \mathcal{T}_{p} $.
    From Lemma \ref{lem:2R-1D1B-Jacobian-Tz}, we obtain that the set $ \mathcal{T}_{z} $ and implicitly also $ \mathcal{T}_{p} $ is unstable. 
    Hence, if $ p\BR{0} \not\in \mathcal{T}_{p} $, it follows $ p\BR{t} \nrightarrow \mathcal{T}_{p} $. 
    This implies that $ \dot{V}\BR{e} = 0 $ if and only if the trajectories reaches a point $ \widehat{p} \in \mathcal{S}_{p} $.
    In this scenario, $ V\BR{e} = 0 $ and $ \dot{p}_{1} = \dot{p}_{2} = \mathbb{0}_{2} $.  
    This completes the proof. 
\end{IEEEproof}

\vspace{0.5\baselineskip}
Previously, Remark \ref{rem:2R-1D1B-Global-Stability} encapsulates the result for when the desired inter-robot distance is $ d_{12}^{\star} < \widehat{d} $. 
In combination with Theorem \ref{thm:2R-1D1B-Almost-Global-Convergence}, we thus have provided the stability analysis for the (\textbf{1D1B}) setup when the closed-loop system is given by \eqref{eq:2R-1D1B-Closed-Loop}. 

We can now provide answers to the three questions raised in Section \ref{subsec:Problem-Setup}. 
For the the (\textbf{1D1B}) setup, we have shown that equilibrium configurations consist of only the correct inter-robot relative positions $ z_{12}^{\star} $ and we have (almost) global convergence towards $ z_{12}^{\star} $. 
Also, there exist moving configurations $ z_{12_{\text{M}}} $. 
However, these are locally not attractive. 
Moreover, as mentioned in Remark \ref{rem:2R-1D1B-Delay-Tp}, the occurrence of moving configurations can be postponed when the gain ratio $ R_{\text{bd}} $ is increased. 

\section{THE (\textbf{1D2B}) ROBOT SETUP} \label{sec:The-1D2B-Setup}
In the previous section, we have analyzed the case of two robots in the (\textbf{1D1B}) setup. 
One observation which stands out is the fact that robots may move with a constant velocity $ w $ when they are initialized at specific points in the plane.

In this section, we consider the case of three robots with the specific partition $ \mathcal{D} = \CBR{1} $ and $ \mathcal{B} = \CBR{2, \, 3} $; the (\textbf{1D2B}) setup depicted in Fig. \ref{fig:Robot-Topology}.
The neighbor sets are found to be $ \mathcal{N}_{1} = \CBR{2, \, 3} $ and $ \mathcal{N}_{2} = \mathcal{N}_{3} = \CBR{1} $. 
Utilizing gradient-based control laws for each distance or bearing task, we obtain the following closed-loop dynamics 
\begin{equation} \label{eq:3R-1D2B-Closed-Loop} 
    \begin{aligned}
        \begin{bmatrix}
            \dot{p}_{1}
            \\
            \dot{p}_{2}
            \\
            \dot{p}_{3}
        \end{bmatrix}
        =
        \begin{bmatrix}
            K_{\text{d}} e_{12\text{d}} z_{12} + K_{\text{d}} e_{13\text{d}} z_{13}
            \\
            K_{\text{b}} e_{21\text{b}}
            \\
            K_{\text{b}} e_{31\text{b}}
        \end{bmatrix}
        ,
    \end{aligned}
\end{equation}
where we assume the distance robot \texttt{R1} has the control gain $ K_{\text{d}} > 0 $ and the bearing robots \texttt{R2} and \texttt{R3} the common control gain $ K_{\text{b}} > 0 $.
In the (\textbf{1D1B}) setup, we observe the link dynamics is also of interest. 
Therefore, in the current setup, we define the relative position vector $ z = {\begin{bmatrix} z_{12}^{\top} & z_{13}^{\top} \end{bmatrix}}^{\top} \in \RN^{4} $.
We have $ z = \widetilde{H}p $ with the graph incidence matrix $ H = \begin{bmatrix} -1 & 1 & 0 \\ -1 & 0 & 1 \end{bmatrix} $ and $ \widetilde{H} = H \otimes I_{2} $.
The link dynamics is then $ \dot{z} = \widetilde{H}\dot{p} $.
In expanded form, we write 
\begin{equation} \label{eq:3R-1D2B-Link-Dynamics}
    \begin{aligned}
        \begin{bmatrix}
            \dot{z}_{12}
            \\
            \dot{z}_{13}
        \end{bmatrix}
        =
        -
        \begin{bmatrix}
            K_{\text{b}} e_{12\text{b}} + K_{\text{d}} e_{12\text{d}} z_{12} + K_{\text{d}} e_{13\text{d}} z_{13}
            \\
            K_{\text{b}} e_{13\text{b}} + K_{\text{d}} e_{12\text{d}} z_{12} + K_{\text{d}} e_{13\text{d}} z_{13}
        \end{bmatrix}
        .
    \end{aligned}
\end{equation}
For a triangle, it holds that $ z_{12} + z_{23} - z_{13} = \mathbb{0}_{2} $. 
Hence, the dynamics related to the `invisible' link $ z_{23} $ evaluates to $ \dot{z}_{23} = - K_{\text{b}} \BR{e_{13\text{b}} - e_{12\text{b}}} $.

In the following subsections, we analyze the closed-loop formation system \eqref{eq:3R-1D2B-Closed-Loop}, following similar steps as we have done for the (\textbf{1D1B}) setup.

\subsection{Equilibrium configurations} \label{subsec:Equilibrium-Configurations}
The following result on the equilibrium configurations is obtained. 

\vspace{0.5\baselineskip}
\begin{proposition}[\emph{(\textbf{1D2B})} Equilibrium Configurations] \label{prop:3R-1D2B-Equilibrium-Configuations}
    The equilibrium configurations corresponding to the closed-loop formation system \eqref{eq:3R-1D2B-Closed-Loop} belong to the set 
    \begin{equation} \label{eq:3R-1D2B-Equilibrium-Set}
        \begin{aligned}
            \mathcal{S}_{p} = \CBR{p \in \RN^{6} \, \rvert \, e = \mathbb{0}_{6}}
            ,
        \end{aligned}
    \end{equation}    
    where $ e = {\begin{bmatrix} e_{12\text{d}} & e_{12\text{d}} & e_{12\text{b}}^{\top} & e_{13\text{b}}^{\top} \end{bmatrix}}^{\top} \in \RN^{6} $.
\end{proposition}

\vspace{0.5\baselineskip}
\begin{IEEEproof}
    Setting the left hand side (LHS) of each equation in \eqref{eq:3R-1D2B-Closed-Loop} to the zero vector, we immediately obtain that at the equilibrium configurations, the bearing constraints for robots \texttt{R2} and \texttt{R3} are satisfied since $ e_{21\text{b}} = - e_{12\text{b}} = \mathbb{0}_{2} $ and $ e_{31\text{b}} = - e_{13\text{b}} =  \mathbb{0}_{2} $. 
    This implies that $ d_{21} = d_{12} \neq  0 $ and $ d_{31} = d_{13} \neq 0 $.
    It remains to solve for $ \dot{p}_{1} = \mathbb{0}_{2} $.
    With the gathered insights, we obtain $ e_{12\text{d}} d_{12} g_{12}^{\star} = - e_{13\text{d}} d_{13} g_{13}^{\star} $.
    Since $ g_{12}^{\star} \neq \pm g_{13}^{\star} $ (the robots are co-linear when $ g_{12} = \pm g_{13} $), the only way to satisfy the expression is when $ e_{12\text{d}} d_{12} = 0 \, \wedge \,  e_{13\text{d}} d_{13} = 0 $.
    Because $ d_{12} \neq 0 $ and also $ d_{13} \neq 0 $, we require $ e_{12\text{d}} = 0 \, \wedge \, e_{13\text{d}} = 0 $ to hold. 
    This completes the proof.
\end{IEEEproof}

\subsection{Moving configurations}
During the analysis of the (\textbf{1D1B}) setup, we observed that robots may move with a common velocity $ w $ while the predefined constraints are not met. 
Recall that $ \dot{z} = \widetilde{H} \dot{p} $ and $ \SPAN{\mathbb{1}_{3}} \subseteq \NULL{H} $.
Hence, for the (\textbf{1D2B}) setup, we aim to identify the conditions such that the formation may move with the common velocity $ w $. 

\vspace{0.5\baselineskip}
\begin{proposition}[\emph{(\textbf{1D2B})} Moving Configurations] \label{prop:3R-1D2B-Moving-Configurations}
    The closed-loop formation system \eqref{eq:3R-1D2B-Closed-Loop} moves with a constant velocity $ w = K_{\text{b}} \BR{g_{12}^{\star} + g_{13}^{\star}} $ when the error vector $ e $ is of the form
    \begin{equation} \label{eq:3R-1D2B-Moving-Conditions}
        \begin{aligned}
            \begin{bmatrix}
                e_{12\text{d}}
                \\
                e_{13\text{d}}
                \\
                e_{12\text{b}}
                \\
                e_{13\text{b}}
            \end{bmatrix}
            =
            -
            \begin{bmatrix}
                \frac{1}{d_{12}} R_{\text{bd}}
                \\
                \frac{1}{d_{13}} R_{\text{bd}}
                \\
                \BR{g_{12}^{\star} + g_{13}^{\star}}
                \\
                \BR{g_{12}^{\star} + g_{13}^{\star}}
            \end{bmatrix}
            .
        \end{aligned}
    \end{equation}
\end{proposition}

\vspace{0.5\baselineskip}
\begin{IEEEproof}
    First, we solve for $ \dot{z} = \mathbb{0}_{4} $. 
    Since $ \dot{z}_{12} = \mathbb{0}_{2} = \dot{z}_{13} $, it follows $ \dot{z}_{23} = \mathbb{0}_{2} $.
    This expression evaluates to $ g_{12} - g_{13} = g_{12}^{\star} - g_{13}^{\star} $.
    Define $ b_{\text{diff}} = g_{12} - g_{13} $ and let $ \angle g_{12} = \alpha $ be the angle enclosed by $ g_{12} $ and the positive $ x $-axis of $ \Sigma^{\text{g}} $.
    Similarly, let $ \angle g_{13} = \beta $.  
    We can rewrite $ b_{\text{diff}} $ as 
    \begin{equation} \label{eq:3R-1D2B-g_diff}
        \begin{aligned}
            b_{\text{diff}}
                & = g_{12} - g_{13}
                = 
                \begin{bmatrix}
                    \cos \alpha 
                    \\
                    \sin \alpha
                \end{bmatrix}
                -
                \begin{bmatrix}
                    \cos \beta 
                    \\
                    \sin \beta 
                \end{bmatrix}
                \\
                & = 
                2 \cos \BR{\frac{\alpha - \beta_{\pi}}{2}}
                \begin{bmatrix}
                    \cos \BR{\frac{\alpha + \beta_{\pi}}{2}} 
                    \\
                    \sin \BR{\frac{\alpha + \beta_{\pi}}{2}}
                \end{bmatrix}
                ,
        \end{aligned}
    \end{equation}
where $ \beta_{\pi} = \beta + \pi \: \BR{\text{mod } 2\pi}$.
The expression $ \dot{z}_{23} = \mathbb{0}_{2} $ can be transformed to the following set of constraints on the angles, namely 
\begin{equation} \label{eq:3R-1D2B-alpha-beta-correct}
    \begin{aligned}
        \begin{cases}
            \alpha + \beta_{\pi}
                = \alpha^{\star} + \beta_{\pi}^{\star} 
                \\
            \alpha - \beta_{\pi}
                = \alpha^{\star} - \beta_{\pi}^{\star} 
        \end{cases}
        \iff 
        \begin{cases}
            \alpha 
                = \alpha^{\star} 
                \\
            \beta 
                = \beta^{\star}        
        \end{cases}  
    \end{aligned}
\end{equation}
and
\begin{equation} \label{eq:3R-1D2B-alpha-beta-incorrect}
    \begin{aligned}
        \begin{cases}
            \hfill \alpha + \beta_{\pi} 
                = \alpha^{\star} + \beta_{\pi}^{\star}   
                \\
            - \BR{\alpha - \beta_{\pi}}
                = \alpha^{\star} - \beta_{\pi}^{\star}
        \end{cases}
        \iff 
        \begin{cases}
            \alpha 
                = \beta^{\star} + \pi
                \\
            \beta 
                = \alpha^{\star} - \pi
        \end{cases}
        .
    \end{aligned}
\end{equation}
From \eqref{eq:3R-1D2B-alpha-beta-correct}, we obtain $ g_{12} = g_{12}^{\star} $ and $ g_{13} = g_{13}^{\star} $, corresponding to the equilibrium configurations in $ \mathcal{S}_{p} $ while the solution in \eqref{eq:3R-1D2B-alpha-beta-incorrect} corresponds to $ g_{12} = - g_{13}^{\star} $ and $ g_{13} = - g_{12}^{\star} $.
Subsequently, we obtain $ e_{12\text{b}} = e_{13\text{b}} = - \BR{g_{12}^{\star} + g_{13}^{\star}} $.
Hence, it is sufficient to consider one of the equations in \eqref{eq:3R-1D2B-Link-Dynamics}.
This leads to $ \BR{- K_{\text{d}} e_{13\text{d}} d_{13}} g_{12}^{\star} + \BR{- K_{\text{d}} e_{12\text{d}}d_{12}} g_{13}^{\star} = K_{\text{b}} g_{12}^{\star} + K_{\text{b}} g_{13}^{\star} $.
For it to hold, we require $ - K_{\text{d}} e_{13\text{d}} d_{13} = K_{\text{b}} \iff e_{13\text{d}} = - \frac{1}{d_{13}} R_{\text{bd}} $ and $ - K_{\text{d}} e_{12\text{d}} d_{12} = K_{\text{b}} \iff e_{12\text{d}} = -\frac{1}{d_{12}} R_{\text{bd}} $ with the gain ratio $ R_{\text{bd}} = \frac{K_{\text{b}}}{K_{\text{d}}} $.
Collecting the error constraints, we obtain \eqref{eq:3R-1D2B-Moving-Conditions}. 
By an immediate substitution, we obtain for the dynamics of the bearing robot \texttt{R2}, $ \dot{p}_{2} = K_{\text{b}} \BR{g_{12}^{\star} + g_{13}^{\star}} \eqqcolon w $. 
This completes the proof.
\end{IEEEproof}

\vspace{0.5\baselineskip}
\begin{remark}
    The signed area for a triangle, introduced as a constraint for resolving flip and flex ambiguities in \cite{Anderson2008, Liu2020}, can be obtained using the expression $ S_{\text{A}} = z_{12}^{\top} \left[ \begin{smallmatrix} 0 & 1 \\ -1 & 0 \end{smallmatrix} \right] z_{13} $. 
    The signed area of the desired formation shape evaluates to $ S_{\text{A}}^{\star} = d_{12}^{\star} d_{13}^{\star} g_{12}^{\top \, \star} \left[ \begin{smallmatrix} 0 & 1 \\ -1 & 0 \end{smallmatrix} \right] g_{13}^{\star} $.
    The signed area of the moving formation shape is $ S_{\text{A}_{\text{M}}} = d_{12} d_{13} g_{13}^{\top \, \star} \left[ \begin{smallmatrix} 0 & 1 \\ -1 & 0 \end{smallmatrix} \right] g_{12}^{\star} = - d_{12} d_{13} g_{12}^{\top \, \star} \left[ \begin{smallmatrix} 0 & 1 \\ -1 & 0 \end{smallmatrix} \right] g_{13}^{\star} = - \frac{d_{12} d_{13}}{d_{12}^{\star} d_{13}^{\star}} S_{\text{A}}^{\star} $.
    Since the distance error signals in \eqref{eq:3R-1D2B-Moving-Conditions} are negative, it follows $ d_{ij} < d_{ij}^{\star} $. 
    Hence $ \ABS{S_{\text{A}_{\text{M}}}} < \ABS{S_{\text{A}}} $ and the cyclic ordering of the robots is opposite to that of the desired formation shape. 
\end{remark}

\vspace{0.5\baselineskip}
Following Proposition \ref{prop:3R-1D2B-Moving-Configurations}, a characterization of the moving set $ \mathcal{T}_{p} $ in terms of the error vector $ e $ is 
\begin{equation} \label{eq:3R-1D2B-Moving-Set}
    \begin{aligned}
        \mathcal{T}_{p} = \CBR{p \in \RN^{6} \, \rvert \, e \text{ satisfies } \eqref{eq:3R-1D2B-Moving-Conditions}}
        .
    \end{aligned}
\end{equation}
An equivalent characterization of $ \mathcal{T}_{p} $ can be provided in terms of the inter-robot relative position vectors $ z_{12_\text{M}} $ and $ z_{13_{\text{M}}} $ where subscript $ _\text{M} $ refers to ``moving''.
In fact, the inter-robot bearing vectors $ g_{12_{\text{M}}} $ between \texttt{R1} and \texttt{R2} and $ g_{13_{\text{M}}} $ between robots \texttt{R1} and \texttt{R3} is known from the proof of Proposition \ref{prop:3R-1D2B-Moving-Configurations}. 
It remains to obtain feasible values for the inter-robot distances $ d_{12_{\text{M}}} $ and $ d_{13_{\text{M}}} $.
To this end, we find the roots satisfying the expressions for the distance error signals $ e_{12\text{d}} $ and $ e_{13\text{d}} $ in \eqref{eq:3R-1D2B-Moving-Conditions}. 
Expanding the expressions leads to the following cubic equation
\begin{equation} \label{eq:3R-1D2B-Cubic-equation}
    \begin{aligned}
        d_{ij}^{3} - \BR{d_{ij}^{\star}}^{2} d_{ij} + R_{\text{bd}} = 0, \, ij \in \CBR{12, \, 13}
        ,
    \end{aligned}
\end{equation}
similar to the (\textbf{1D1B}) setup. 
Compared with \eqref{eq:Prel-Reduced-Cubic}, we now have $ c = - \BR{d_{ij}^{\star}}^{2} < 0 $ and $ d = R_{\text{bd}} > 0 $.
In the (\textbf{1D1B}) setup, we have $ d = 2 R_{\text{bd}} $.
Following the same steps as was done for \eqref{eq:2R-1D1B-Cubic-Equation-d12}, we obtain that the discriminant corresponding to \eqref{eq:3R-1D2B-Cubic-equation} is $ \Delta = 4 \BR{d_{ij}^{\star}}^{6} - 27 R_{\text{bd}}^{2} $ and the threshold value for the desired distance $ d_{ij}^{\star} $ such that positive roots exist is $ \widehat{d} = \sqrt{3} \sqrt[3]{\frac{R_{\text{bd}}}{2}} \approx 1.3747 \, \sqrt[3]{R_{\text{bd}}} $.
Similar to the (\textbf{1D1B}) setup, we conclude that if one of (or both) the desired distances $ d_{12}^{\star} $ or (and) $ d_{13}^{\star} $ has (have) a value less than $ \widehat{d} $, then no feasible value for $ d_{12} $ or (and) $ d_{13} $ satisfies $ e_{12\text{d}} d_{12} = - R_{\text{bd}} $ or (and) $ e_{13\text{d}} d_{13} = - R_{\text{bd}} $, implying the in-feasibility of moving formations.
We conclude the set $ \mathcal{S}_{p}$ containing equilibrium configurations with either one of (or both) $d_{12}^{\star}<\widehat{d}$ or $d_{13}^{\star}<\widehat{d}$ is  asymptotically stable. The asymptotic stability property follows the same arguments as that in the following subsection (for the case when $d_{12}^{\star}\geq\widehat{d}$ and $d_{13}^{\star}\geq\widehat{d}$), and therefore it is omitted.

When the desired distance $ d_{ij}^{\star} $ satisfies $ d_{ij}^{\star} \geq \widehat{d} $, we obtain feasible distances $ d_{ij} $ to \eqref{eq:3R-1D2B-Cubic-equation} are given by Lemma \ref{lem:Prel-Reduced-Cubic-Roots-Positive} with the values $ r_{v} = \sqrt{\frac{\BR{d_{ij}^{\star}}^{6}}{27}} $ and $ \varphi_{v} = \tan^{-1} \BR{-2 R_{\text{bd}}^{-1} \sqrt{-R}} $.
Since we have two desired distances $ d_{12}^{\star} $ and $ d_{13}^{\star} $ and we have either one or two feasible value(s) $ d_{ij} $ to the cubic equation \eqref{eq:3R-1D2B-Cubic-equation}, it follows that different feasible combinations $ \BR{d_{12_{\text{M}}}, \, d_{13_{\text{M}}}} $ exist. 
In Table \ref{tab:3R-1D2B-Feasible-Combinations}, we summarize the number of feasible combinations for the different scenarios.
We conclude the set $ \mathcal{T}_{p} \neq \emptyset $ when the additional constraints $ d_{12}^{\star} \geq \widehat{d} $ and $ d_{13}^{\star} \geq \widehat{d} $ are satisfied. 
\begin{table}[!tb]
    \centering
    \caption{Number of feasible combinations $ \BR{d_{12_{\text{M}}}, \, d_{13_{\text{M}}}} $ depending on desired distances $ d_{12}^{\star} $ and $ d_{13}^{\star} $}
    \label{tab:3R-1D2B-Feasible-Combinations}
    \begin{tabularx}{0.5\linewidth}{ccc}
        \toprule 
        & $ d_{13}^{\star} > \widehat{d} $ & $ d_{13}^{\star} = \widehat{d} $
        \\
        \midrule
        $ d_{12}^{\star} > \widehat{d} $ & $ 4 $ & $ 2 $
        \\
        $ d_{12}^{\star} = \widehat{d} $ & $ 2 $ & $ 1 $
        \\
        \bottomrule
    \end{tabularx}
\end{table}

Recall the common velocity $ w = K_{\text{b}} \BR{g_{12}^{\star} + g_{13}^{\star}} $ for the robots in Proposition \ref{prop:3R-1D2B-Moving-Configurations}.
Define $ b_{\text{sum}} = g_{12} + g_{13} $. 
We want to write $ b_{\text{sum}} $ in the form $ b_{\text{sum}} = d_{\text{sum}} \, g_{\text{sum}} $ with $ d_{\text{sum}} $ being the magnitude and $ g_{\text{sum}} $ the orientation of $ b_{\text{sum}} $ relative to the global coordinate frame $ \Sigma^{\text{g}} $.
By the sum-to-product identities for cosine and sine, we obtain
\begin{equation} \label{eq:3R-1D2B-g_sum}
    \begin{aligned}
        b_{\text{sum}}
            & = g_{12} + g_{13}
            =
            \begin{bmatrix}
                \cos \alpha
                \\
                \sin \alpha
            \end{bmatrix}
            +
            \begin{bmatrix}
                \cos \beta
                \\
                \sin \beta
            \end{bmatrix}
            \\
            & =
            2 \cos \BR{\frac{\alpha - \beta}{2}} 
            \begin{bmatrix}
                \cos \BR{\frac{\alpha + \beta}{2}}
                \\
                \sin \BR{\frac{\alpha + \beta}{2}}
                \\                
            \end{bmatrix}
            .
    \end{aligned}
\end{equation}
Depending on the value of the angle difference $ \ABS{\alpha - \beta } $, we have different expressions for $ d_{\text{sum}} $ and $ g_{\text{sum}} $. 
When $ \ABS{\alpha - \beta} < \pi $, we set $ d_{\text{sum}} = 2 \cos \BR{\frac{\ABS{\alpha - \beta}}{2}} $ and $ \angle g_{\text{sum}} = \frac{\alpha + \beta}{2} $ while for $ \ABS{\alpha - \beta} > \pi $, we set $ d_{\text{sum}} = 2 \cos \BR{\pi - \frac{\ABS{\alpha - \beta}}{2}} $ and $ \angle g_{\text{sum}} = \frac{\alpha + \beta}{2} + \pi \; \BR{\text{mod } 2\pi} $. 
Note that $ d_{\text{sum}} \in \BR{0, \, 2} $ for the cases $ \ABS{\alpha - \beta } \gtrless \pi $.
If $ \ABS{\alpha - \beta} = 0^{\degree} $, then $ g_{12} = g_{13} $ and $ b_{\text{sum}} = 2 g_{12} $, and finally, $ \ABS{\alpha - \beta} = \pi $ implies $ g_{12} = - g_{13} $ and hence $ b_{\text{sum}} = \mathbb{0}_{2} $.
Since $ g_{12}^{\star} \neq \pm g_{13}^{\star} $, the last two mentioned cases does not occur; therefore, the magnitude of $ w $ is $ 0 < \norm{w} < 2 K_{\text{b}} $.

\subsection{Local stability analysis of the equilibrium and moving formations}
Assume that the desired distances satisfy $ d_{12}^{\star} \geq \widehat{d} $ and $ d_{13}^{\star} \geq \widehat{d} $. 
In this case, both the equilibrium configurations in \eqref{eq:3R-1D2B-Equilibrium-Set} and moving configurations in \eqref{eq:3R-1D2B-Moving-Set} satisfy $ \dot{z} = \mathbb{0}_{4} $ and are feasible.
We are interested in determining the local stability around these formations. 
To this end, we consider the linearization of the $ z $-dynamics \eqref{eq:3R-1D2B-Link-Dynamics}; this results in the Jacobian matrix $ A \in \RN^{4 \times 4} $ as
\begin{equation} \label{eq:3R-1D2B-Jacobian}
    \begin{aligned}
        A =
        -
        \begin{bmatrix}
            K_{\text{b}} A_{12\text{b}} + K_{\text{d}} A_{12\text{d}} & K_{\text{d}} A_{13\text{d}}
            \\
            K_{\text{d}} A_{12\text{d}} & K_{\text{b}} A_{13\text{b}} + K_{\text{d}} A_{13\text{d}}
        \end{bmatrix}
        ,
    \end{aligned}
\end{equation}
where $ A_{ij\text{d}} = e_{ij\text{d}} I_{2} + 2 z_{ij} z_{ij}^{\top} $ and $ A_{ij\text{b}} = \frac{1}{d_{ij}} \BR{I_{2} - g_{ij} g_{ij}^{\top}} $, $ ij \in \CBR{12, \, 13} $.

We first consider the stability analysis around the equilibrium configurations.

\vspace{0.5\baselineskip}
\begin{theorem} \label{thm:3R-1D2B-Jacobian-Equilibrium}
    Consider a team of three robots arranged in the (\textbf{1D2B}) setup with closed-loop dynamics given by \eqref{eq:3R-1D2B-Closed-Loop}. 
    Assume the desired distances satisfy $ d_{12}^{\star} \geq \widehat{d} $ and $ d_{13}^{\star} \geq \widehat{d} $ with $ \widehat{d} = \sqrt{3} \sqrt[3]{\frac{R_{\text{bd}}}{2}} $ and the bearing vectors satisfy $ g_{12}^{\star} \neq \pm g_{13}^{\star} $.
    Given an initial configuration $ p\BR{0} $ that is close to the desired formation shape, then the robot trajectories asymptotically converge to a point $ \widehat{p} \in \mathcal{S}_{p} $.
\end{theorem}

\vspace{0.5\baselineskip}
\begin{IEEEproof}
    The proof for the local asymptotic stability of the equilibrium configurations in $ \mathcal{S}_{p} $ is shown using Lyapunov's indirect method.
    
    Evaluating the Jacobian matrix \eqref{eq:3R-1D2B-Jacobian} at the equilibrium configurations results in 
    \begin{equation} \label{eq:3R-1D2B-Equilibrium-Jacobian}
        \begin{aligned}
            & A_{\text{E}}
                = - \DIAG{
                \begin{bmatrix}
                    x^{\star} & x^{\star} & p^{\star} & p^{\star} 
                \end{bmatrix}
                }
                -
            \\
            &   
                \begin{bmatrix}
                    \BR{y^{\star} - x^{\star}} a^{2} & \BR{y^{\star} - x^{\star}} ab & q^{\star} c^{2} & q^{\star} cd
                    \\
                    \BR{y^{\star} - x^{\star}} ab & \BR{y^{\star} - x^{\star}} b^{2} & q^{\star}cd & q^{\star}d^{2}
                    \\
                    y^{\star} a^{2} & y^{\star}ab & \BR{q^{\star} - p^{\star}} c^{2} & \BR{q^{\star} - p^{\star}} cd
                    \\
                    y^{\star} ab & y^{\star} b^{2} & \BR{q^{\star} - p^{\star}} cd & \BR{q^{\star} - p^{\star}} d^{2}
                \end{bmatrix}
                ,
        \end{aligned}
    \end{equation}
    where we define the variables
    \begin{equation} \label{eq:3R-1D2B-Jacobian-Variables}
        \begin{aligned}
            x = K_{\text{b}} \frac{1}{d_{12}},
            \, 
            y = 2 K_{\text{d}} d_{12}^{2},
            \,
            p = K_{\text{b}}  \frac{1}{d_{13}},
            \,
            q = 2 K_{\text{d}} d_{13}^{2}
            .
        \end{aligned}
    \end{equation}
    and the matrices 
    \begin{equation} \label{eq:3R-1D2B-Jacobian-Matrices}
        \begin{aligned}
            g_{12}^{\star}g_{12}^{\star \, \top} 
                & = \begin{bmatrix} a^{2} & ab \\ ab & b^{2} \end{bmatrix},
                &
            g_{13}^{\star} g_{13}^{\star \, \top} 
                & = \begin{bmatrix} c^{2} & cd \\ cd & d^{2} \end{bmatrix}
                .
        \end{aligned}
    \end{equation}
    The starred version for $ x, \, p, \, y $, and $ q $ is used here since we have $ d_{12} = d_{12}^{\star} $ and $ d_{13} = d_{13}^{\star} $.
    The characteristic polynomial $ \chi_{\text{E}} \BR{\lambda} $ corresponding to matrix $ A_{\text{E}} $ is obtained as 
    \begin{equation} \label{eq:3R-1D2B-Equilibrium-Characteristic}
        \begin{aligned}
            \chi_{\text{E}} \BR{\lambda} 
                & = \BR{\lambda + x^{\star}} \BR{\lambda + p^{\star}} \dots 
                \\
                & \quad \BR{\lambda^{2} + \BR{y^{\star} + q^{\star}} \lambda + y^{\star}q^{\star} \sin^{2} \theta^{\star}}
                ,
        \end{aligned}
    \end{equation}
    where $ \sin \theta = g_{12}^{\top} \left[ \begin{smallmatrix} 0 & 1 \\ -1 & 0 \end{smallmatrix} \right] g_{13} $.
    The roots of \eqref{eq:3R-1D2B-Equilibrium-Characteristic} are
    \begin{equation} \label{eq:3R-1D2B-Equilibrium-Roots}
        \begin{aligned}
            \lambda_{1} & = - x^{\star}, \quad \lambda_{2} = - p^{\star}, 
            \\
            \lambda_{3, \, 4} & = - \frac{1}{2} \BR{y^{\star} + q^{\star}} \pm \frac{1}{2} \sqrt{\BR{y^{\star} + q^{\star}}^{2} - 4y^{\star}q^{\star} \sin^{2} \theta^{\star}}
            .
        \end{aligned}
    \end{equation}
    It can be verified that $ 0 < 4y^{\star}q^{\star} \sin^{2} \theta^{\star} \leq \BR{y^{\star} + q^{\star}}^{2}  $; therefore, all roots are real.
    Moreover, $ - \frac{1}{2} \BR{y^{\star} + q^{\star}} + \frac{1}{2} \sqrt{\BR{y^{\star} + q^{\star}}^{2} - 4y^{\star}q^{\star} \sin^{2} \theta^{\star}} < 0 $ and we conclude all $ \lambda $s are negative; hence, the matrix $ A_{\text{E}} $ is Hurwitz.
    This implies that the link trajectories asymptotically converge to the desired relative positions, i.e., $ z\BR{t} \rightarrow z^{\star} $ as $ t \rightarrow \infty $. 
    It also means that the robots reach their individual tasks since $ z_{ij}^{\star} = d_{ij}^{\star} g_{ij}^{\star} $, so $ p\BR{t} \rightarrow \mathcal{S}_{p} $ when $ p\BR{0} $ is close to the desired formation shape.
    This completes the proof.
\end{IEEEproof}

\vspace{0.5\baselineskip}
We continue with determining the stability of the moving formations in the set $ \mathcal{T}_{p} $.
Again, the Lyapunov's indirect method is used for this task. 
As a first step, the characteristic polynomial corresponding to the Jacobian matrix of the moving formations is derived. 
Based on the characterization in \eqref{eq:3R-1D2B-Moving-Conditions}, we obtain the sub-matrices $ A_{12\text{d}} = - R_{\text{bd}} \frac{1}{d_{12}} I_{2} + 2 d_{12}^{2} g_{13}^{\star} g_{13}^{\star \, \top} $, $ A_{13\text{d}} = -R_{\text{bd}} \frac{1}{d_{13}} I_{2} + 2 d_{13}^{2} g_{12}^{\star} g_{12}^{\star \, \top} $, $ A_{12\text{b}} = \frac{1}{d_{12}} \BR{I_{2} - g_{13}^{\star} g_{13}^{\star, \top}} $, and $ A_{13\text{b}} = \frac{1}{d_{13}} \BR{I_{2} - g_{12}^{\star} g_{12}^{\star, \top}} $.
Substituting these sub-matrices in \eqref{eq:3R-1D2B-Jacobian} yields the matrix 
\begin{equation} \label{eq:3R-1D2B-Moving-Jacobian}
    A_{\text{M}} =
    -
    \begin{bmatrix}
        \BR{y - x} c^{2} & \BR{y - x} cd & -p + q a^{2} & q ab
        \\
        \BR{y - x} cd & \BR{y - x} d^{2} & q ab & -p + q b^{2}
        \\
        -x + y c^{2} & ycd & \BR{q - p} a^{2} & \BR{q - p} ab
        \\
        ycd & -x + y d^{2} &  \BR{q - p} ab & \BR{q - p} b^{2}        
    \end{bmatrix}
    ,
\end{equation}
where the variables are previously defined in \eqref{eq:3R-1D2B-Jacobian-Variables} and \eqref{eq:3R-1D2B-Jacobian-Matrices}.
The characteristic polynomial $ \chi_{\text{M}} \BR{\lambda} $ corresponding to matrix $ A_{\text{M}} $ is then obtained as the quartic polynomial
\begin{equation} \label{eq:3R-1D2B-Moving-Characteristic}
    \begin{aligned}
        \chi_{\text{M}} \BR{\lambda}
            = \lambda^{4} + c_{1} \lambda^{3} + c_{2} \lambda^{2} + c_{3} \lambda + c_{4}
    \end{aligned}
\end{equation}
with the coefficients
\begin{equation} \label{eq:3R-1D2B-Moving-Characteristic-Coefficients}
    \begin{aligned}
        c_{1} 
            & = \BR{y - x} + \BR{q - p},
        \\
        c_{2} 
            & = qy\sin^{2} \theta^{\star} - px, 
        \\
        c_{3} 
            & = x\BR{y - x} \BR{q \sin^{2} \theta^{\star} - p} + p\BR{q - p} \BR{y \sin^{2} \theta^{\star} - x},
        \\
        c_{4} 
            & = p x\BR{y - x} \BR{q - p} \sin^{2} \theta^{\star}.
    \end{aligned}
\end{equation}

Recall from Table \ref{tab:3R-1D2B-Feasible-Combinations} that depending on the value of the desired distances $ d_{12}^{\star} $ and $ d_{13}^{\star} $, we can obtain more than one feasible combination $ \BR{d_{12_{\text{M}}}, \, d_{13_{\text{M}}}} $ for the moving configurations. 
Under certain conditions, we have the following result on the eigenvalues of the matrix $ A_{\text{M}} $.

\vspace{0.5\baselineskip}
\begin{lemma} \label{lem:3R-1D2B-Moving-Formation-Negative-Eigenvalue}
    Assume the desired distances satisfy $ d_{12}^{\star} > \widehat{d} $ and $ d_{13}^{\star} > \widehat{d} $ and the desired bearing vectors are not perpendicular, i.e. $ g_{12}^{\star} \not\perp g_{13}^{\star} $. 
    Consider the feasible combination in which the distances are of the form $ d_{12_{\text{M}}} = y_{p_{1}}\BR{d_{12}^{\star}} $ and $ d_{13_{\text{M}}} = y_{p_{1}}\BR{d_{13}^{\star}} $ in Lemma \ref{lem:Prel-Reduced-Cubic-Roots-Positive}.
    Then all eigenvalues of the matrix $ A_{\text{M}} $ has negative real part if the inequality
    \begin{equation} \label{eq:3R-1D2B-Moving-Formation-Cosine}
        \cos^{2} \theta^{\star} < \frac{\BR{mq - ny}^{2} + mn\BR{m + n}\BR{x + p}}{\BR{m^{2}q + n^{2}y}\BR{mqx + nyp}}
        ,
    \end{equation}
    holds, 
    where the variables $ x $, $ y $, $ p $, and $ q $ are as defined in \eqref{eq:3R-1D2B-Jacobian-Variables}, and we define $ m = y - x $ and $ n = q - p $.
\end{lemma}

\vspace{0.5\baselineskip}
\begin{IEEEproof}
    Assuming the bearing vectors are not perpendicular, we obtain that $ 0 < \sin^{2} \theta^{\star} < 1 $.
    Also, since $ d_{12_{\text{M}}} = y_{p_{1}}\BR{d_{12}^{\star}} $ and $ d_{13_{\text{M}}} = y_{p_{1}}\BR{d_{13}^{\star}} $ and $ d_{12}^{\star} > \widehat{d} $ and $ d_{13}^{\star} > \widehat{d} $, we can verify that $ m > 0 $ and $ n > 0 $ by applying Proposition \ref{prop:App-Expression-h}.
    We employ the Routh-Hurwitz stability criterion to show the desired result provided \eqref{eq:3R-1D2B-Moving-Formation-Cosine} holds. 
    To this end, the Routh-Hurwitz table is formed. 
    The first column of the Routh-Hurwitz table, which is the column of interest, contains the following values 
    \begin{equation} \label{eq:3R-1D2B-RH-Criterion}
        \begin{bmatrix}
            1 & c_{1} & \frac{c_{1}c_{2} - c_{3}}{c_{1}} & \frac{\BR{c_{1}c_{2} - c_{3}} c_{3} - c_{1}^{2} c_{4}}{\BR{c_{1}c_{2} - c_{3}}} & c_{4}
        \end{bmatrix}
        .
    \end{equation}
    For all roots $ \lambda $ to have negative real parts, all values in \eqref{eq:3R-1D2B-RH-Criterion} need to be positive.
    With $ m > 0 $ and $ n > 0 $, the coefficients $ c_{1} $ and $ c_{4} $ are positive. 
    It remains to show the third and fourth entry in \eqref{eq:3R-1D2B-RH-Criterion} is positive. 
    In fact, it is sufficient to show the numerators are both positive.
    The numerator $ c_{1} c_{2} - c_{3} $ evaluates to 
    \begin{equation} \label{eq:3R-1D2B-RH-Criterion-Case-21}
        \begin{aligned}
            & c_{1} c_{2} - c_{3} 
            \\
                & = \BR{m + n} \BR{qy\sin^{2} \theta^{\star} - px} \dots 
                \\
                & \quad - mx\BR{q \sin^{2} \theta^{\star} - p} - np \BR{y \sin^{2} \theta^{\star} - x}
                \\
                & = \sin^{2} \theta^{\star} \BR{m^{2} q + n^{2} y} > 0
                .
        \end{aligned}
    \end{equation}   
    The numerator $ \BR{c_{1}c_{2} - c_{3}} c_{3} - c_{1}^{2} c_{4} $ evaluates to the expression \eqref{eq:3R-1D2B-RH-Criterion-Case-22}.
    \begin{figure*}[!tb]
        \normalsize
        \setcounter{mytempeqncnt}{\value{equation}}
        \begin{equation} \label{eq:3R-1D2B-RH-Criterion-Case-22}
            \begin{aligned}
                & \BR{c_{1}c_{2} - c_{3}} c_{3} - c_{1}^{2} c_{4}
                \\
                & = \BR{\sin^{2} \theta^{\star} \BR{m^{2} q + n^{2} y}} \BR{mx \BR{q \sin^{2} \theta^{\star} - p} + np \BR{y \sin^{2} \theta^{\star} - x}} - \BR{m + n}^{2} \BR{xm pn \sin^{2} \theta^{\star}}
                \\
                & = \sin^{2} \theta^{\star} \BR{\BR{m^{2} q + n^{2} y} \BR{mx \BR{q \BR{\sin^{2} \theta^{\star} - 1} + n} + np \BR{y \BR{\sin^{2} \theta^{\star} - 1} + m}} - \BR{m + n}^{2} xm pn}
                \\
                & = \sin^{2} \theta^{\star} \BR{\BR{\BR{mq - ny}^{2} + mn\BR{m + n} \BR{x + p}} mn - \BR{m^{2} q + n^{2} y} \BR{mxq + npy} \cos^{2} \theta^{\star}}
            \end{aligned}
        \end{equation}
        \hrulefill
        \vspace*{1pt}
    \end{figure*}
    Provided \eqref{eq:3R-1D2B-Moving-Formation-Cosine} holds, it follows the numerator $ \BR{c_{1}c_{2} - c_{3}} c_{3} - c_{1}^{2} c_{4} > 0 $.
    Since the entries in \eqref{eq:3R-1D2B-RH-Criterion} are all positive, we conclude all eigenvalues of the matrix $ A_{\text{M}} $ have negative real parts. 
    This completes the proof.
\end{IEEEproof}

\vspace{0.5\baselineskip}
\begin{remark}
    The implication of Lemma \ref{lem:3R-1D2B-Moving-Formation-Negative-Eigenvalue} is that under certain conditions on the distance and bearing constraints, a subset of the moving set $ \mathcal{T}_{p} $ is locally asymptotically stable. 
    Hence, initializing the robots close to the conditions for the moving formation is not desirable. 
    An illustration of this behavior is provided in Fig. \ref{fig:3R-1D2B-Moving}.
\end{remark}

\vspace{0.5\baselineskip}
Lemma \ref{lem:3R-1D2B-Moving-Formation-Negative-Eigenvalue} also holds when the desired bearing vectors are perpendicular, i.e. $ g_{12}^{\star} \perp g_{13}^{\star} \iff \sin^{2} \theta^{\star} = 1 $. 
In this case, the coefficients in \eqref{eq:3R-1D2B-Moving-Characteristic-Coefficients} and also all entries in \eqref{eq:3R-1D2B-RH-Criterion} are positive; therefore, the matrix $ A_{\text{M}} $ will only have eigenvalues with negative real parts.

A full characterization of the remaining cases can be found in Appendix \ref{subsec:App-3R-1D2B-Moving-Formation}. 
In almost all cases, the matrix $ A_{\text{M}} $ contains at least a root with positive real part and hence, it is not Hurwitz.

\section{THE (\textbf{1B2D}) ROBOT SETUP} \label{sec:The-1B2D-Setup}
In this section, the formation setup with one bearing and two distance robots (\textbf{1B2D}) is considered. 
Without loss of generality, we assume robot \texttt{R1} is the bearing robot while robots \texttt{R2} and \texttt{R3} are the distance robots. 
The rightmost graph in Fig. \ref{fig:Robot-Topology} depicts the interconnection structure for this setup. 
Based on the interconnection structure, the closed-loop dynamics is obtained as 
\begin{equation} \label{eq:3R-1B2D-Closed-Loop}
    \begin{aligned}
        \begin{bmatrix}
            \dot{p}_{1}
            \\
            \dot{p}_{2}
            \\
            \dot{p}_{3}
        \end{bmatrix}
        =
        \begin{bmatrix}
            K_{\text{b}} e_{12\text{b}} + K_{\text{b}} e_{13\text{b}}
            \\
            K_{\text{d}} e_{21\text{d}} z_{21}
            \\
            K_{\text{d}} e_{31\text{d}} z_{31}
        \end{bmatrix}
        .
    \end{aligned}
\end{equation}
The corresponding link dynamics evaluates to
\begin{equation} \label{eq:3R-1B2D-Link-Dynamics}
    \begin{aligned}
        \begin{bmatrix}
            \dot{z}_{12}
            \\
            \dot{z}_{13}
        \end{bmatrix}
        =
        -
        \begin{bmatrix}
            K_{\text{d}} e_{12\text{d}} z_{12} + K_{\text{b}} e_{12\text{b}} + K_{\text{b}} e_{13\text{b}}
            \\
            K_{\text{d}} e_{13\text{d}} z_{13} + K_{\text{b}} e_{12\text{b}} + K_{\text{b}} e_{13\text{b}}
        \end{bmatrix}
    \end{aligned}
    .
\end{equation}
Also, the dynamics of the `invisible' link $ z_{23} $ is found to be $ \dot{z}_{23} = - K_{\text{d}} \BR{e_{13\text{d}}z_{13} - e_{12\text{d}} z_{12}} $.
In the following, we follow similar steps as in Sections \ref{sec:The-1D1B-Setup} and \ref{sec:The-1D2B-Setup} for the analysis of the closed-loop formation system \eqref{eq:3R-1B2D-Closed-Loop} focusing on equilibrium configurations, possible moving formations, and their (local) stability analysis. 

\subsection{Equilibrium configurations}
When we consider the equilibrium conditions with $ \dot{p} = \mathbb{0}_{6} $, we have the following result.

\vspace{0.5\baselineskip}
\begin{proposition}[\emph{(\textbf{1B2D})} Equilibrium Configurations:] \label{prop:3R-1B2D-Equilibrium-Configurations}
    The equilibrium configurations corresponding to the closed-loop formation system \eqref{eq:3R-1B2D-Closed-Loop} 
    belong to $\mathcal{S}_{p}^{\text{correct}}\cup \mathcal{S}_{p}^{\text{flipped}}$, where 
    \begin{equation} \label{eq:3R-1B2D-Equilibrium-Sets}
        \begin{aligned}
            \mathcal{S}_{p}^{\text{correct}}
                & = \CBR{p \in \RN^{6} \, \rvert \, e = \mathbb{0}_{6}}
                ,
                \\
            \mathcal{S}_{p}^{\text{flipped}}
                & = \CBR{p \in \RN^{6} \, \rvert \, e = {\begin{bmatrix} 0 & 0 & - b_{\text{diff}}^{\star \, \top} & b_{\text{diff}}^{\star \, \top} \end{bmatrix}}^{\top}}
                ,
        \end{aligned}
    \end{equation}
    with $ e = {\begin{bmatrix} e_{12\text{d}} & e_{12\text{d}} & e_{12\text{b}}^{\top} & e_{13\text{b}}^{\top} \end{bmatrix}}^{\top} \in \RN^{6} $ and $ b_{\text{diff}}^{\star} = g_{12}^{\star} - g_{13}^{\star} $.
\end{proposition}

\vspace{0.5\baselineskip}
\begin{IEEEproof}
    Setting the LHS of each equation of the closed-loop dynamics \eqref{eq:3R-1B2D-Closed-Loop} to the zero vector, we obtain for robot \texttt{R2} that $ - K_{\text{d}} e_{12\text{d}} z_{12} = \mathbb{0}_{2} \iff e_{12\text{d}} = 0 \vee z_{12} = \mathbb{0}_{2} $ and similarly, we have $ - K_{\text{d}} e_{13\text{d}} z_{13} = \mathbb{0}_{2} \iff e_{13\text{d}} = 0 \vee z_{13} = \mathbb{0}_{2} $ for \texttt{R3}. 
    The expression for \texttt{R1} evaluates to $ g_{12} + g_{13} = g_{12}^{\star} + g_{13}^{\star} $.
    Defining $ \angle g_{12} = \alpha $, $ \angle g_{13} = \beta $ as before, and recalling the RHS of \eqref{eq:3R-1D2B-g_sum},
    we can write the following set of constraints on the angles, namely
    \begin{equation} \label{eq:3R-1B2D-alpha-beta-correct}
        \begin{aligned}
            \begin{cases}
                \alpha + \beta 
                    = \alpha^{\star} + \beta^{\star}
                    \\
                \alpha - \beta
                    = \alpha^{\star} - \beta^{\star}
            \end{cases}
            \iff 
            \begin{cases}
                \alpha 
                    = \alpha^{\star} 
                    \\
                \beta 
                    = \beta^{\star}        
            \end{cases}  
        \end{aligned}
    \end{equation}
    and
    \begin{equation} \label{eq:3R-1B2D-alpha-beta-incorrect}
        \begin{aligned}
            \begin{cases}
                \hfill \alpha + \beta 
                    = \alpha^{\star} + \beta^{\star}
                    \\
                - \BR{\alpha - \beta}
                    = \alpha^{\star} - \beta^{\star}
            \end{cases}
            \iff 
            \begin{cases}
                \alpha 
                    = \beta^{\star}
                    \\
                \beta  
                    = \alpha^{\star}
            \end{cases}
            .
        \end{aligned}
    \end{equation}
    Equation \eqref{eq:3R-1B2D-alpha-beta-correct} translates to $ g_{12} = g_{12}^{\star} $ and $ g_{13} = g_{13}^{\star} $ implying robot \texttt{R1} satisfies its bearing tasks while \eqref{eq:3R-1B2D-alpha-beta-incorrect} translates to the \textit{flipped} formation shape with bearings satisfying $ g_{12} = g_{13}^{\star} $ and $ g_{13} = g_{12}^{\star} $.
    It follows the bearing error signals are $ e_{12\text{b}} = - e_{13\text{b}} = - b_{\text{diff}}^{\star} $.
    With both $ g_{12} $ and $ g_{13} $ defined, we obtain that $ d_{12} \neq 0 $ and $ d_{13} \neq 0 $. 
    Therefore, $ z_{12} = \mathbb{0}_{2} $ and $ z_{13} = \mathbb{0}_{2} $ are both not feasible. 
    Robots \texttt{R2} and \texttt{R3} will stop moving when $ e_{12\text{d}} = 0 $ and $ e_{13\text{d}} = 0 $ holds, respectively, i.e., when they accomplished their individual distance task irrespective of \texttt{R1}.
    This completes the proof.
\end{IEEEproof}

\vspace{0.5\baselineskip}
It can be verified that the signed area of the flipped formation satisfies $ S_{\text{A}_{\text{F}}} = - S_{\text{A}}^{\star} $.

\subsection{Moving configurations}
For the moving formations, we set the link dynamics to the zero vector and obtain the following result. 

\vspace{0.5\baselineskip}
\begin{proposition}[\emph{(\textbf{1B2D})} Moving Formations] \label{prop:3R-1B2D-Moving-Configurations}
    The moving formations for the (\textbf{1B2D}) setup occur when the robots are co-linear, i.e., $ g_{12} = \pm g_{13} $ and oriented in the direction of $ b_{\text{sum}}^{\star} = \BR{g_{12}^{\star} + g_{13}^{\star}} $.
\end{proposition}

\vspace{0.5\baselineskip}
\begin{IEEEproof}
    The expression for the link $ z_{12} $ and $ z_{13} $ in \eqref{eq:3R-1B2D-Link-Dynamics} evaluates to 
    \begin{equation} \label{eq:3R-1B2D-Link-Dynamics-Zero}
        \begin{aligned}
            \BR{K_{\text{d}} e_{12\text{d}} d_{12} + K_{\text{b}}} g_{12} + K_{\text{b}} g_{13}
                & = K_{\text{b}} b_{\text{sum}}^{\star}
                \\
            K_{\text{b}} g_{12} + \BR{K_{\text{d}} e_{13\text{d}} d_{13} + K_{\text{b}}} g_{13}
                & = K_{\text{b}} b_{\text{sum}}^{\star}
                .
        \end{aligned}
    \end{equation}
    Solving for $ \dot{z}_{23} = \mathbb{0}_{2} $, we obtain $ e_{12\text{d}} d_{12} g_{12} = e_{13\text{d}} d_{13} g_{13} $. 
    Two vectors are equal when they have the same magnitude and direction or opposite signs in both the magnitude and direction.
    Hence we distinguish the case $ g_{12} = g_{13} \wedge e_{12\text{d}} d_{12} = e_{13\text{d}} d_{13} $ and 
    $ g_{12} = - g_{13} \wedge e_{12\text{d}} d_{12} = - e_{13\text{d}} d_{13} $.
    Since $ g_{12} = \pm g_{13} $, we conclude the robots are co-linear.
    Substituting this in \eqref{eq:3R-1B2D-Link-Dynamics-Zero}, we obtain expressions of the form $ h \, g_{12} = K_{\text{b}} d_{\text{sum}}^{\star} g_{\text{sum}}^{\star} $ where $ h = K_{\text{d}} e_{12\text{d}} d_{12} + 2 K_{\text{b}} $ when $ g_{12} = g_{13} $ and $ h = K_{\text{d}} e_{12\text{d}} d_{12} $ when $ g_{12} = - g_{13} $. 
    From this, we infer $ g_{12} = \pm g_{\text{sum}}^{\star} $, implying the orientation of the formation is in the direction of $ b_{\text{sum}}^{\star} $.
    This completes the proof.
\end{IEEEproof}

\begin{figure}[!tb]
    \centering
    \includegraphics[width=0.95\linewidth]{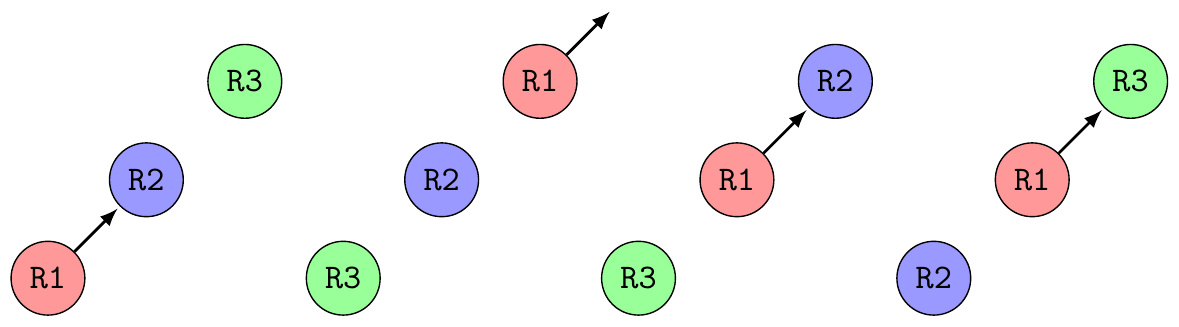}
    \caption{Different robot ordering for the moving configurations in the (\textbf{1B2D}) setup; the black arrow is the bearing vector $ g_{\text{sum}}^{\star} $. From left to right, we have the ordering I to IV. Despite the different colors, both \texttt{R2} and \texttt{R3} are distance robots.}
    \label{fig:3R-1B2D-Colinear}
\end{figure}

\vspace{0.5\baselineskip}
In light of Proposition \ref{prop:3R-1B2D-Moving-Configurations}, we can obtain \textit{four} different ordering of the robots, as depicted in Fig. \ref{fig:3R-1B2D-Colinear}.
To provide a full characterization of the moving configurations, it remains to obtain the inter-robot distances for the different ordering. 
We first derive expressions for the distance error signals corresponding to the different ordering from the general expression $ h \, g_{12} = K_{\text{b}} d_{\text{sum}}^{\star} g_{\text{sum}}^{\star} $. 
Define $ e_{12\text{d}} = \frac{s}{d_{12}} R_{\text{bd}} $ and $ e_{13\text{d}} = \frac{t}{d_{13}} R_{\text{bd}} $. 
Table \ref{tab:3R-1B2D-Distance-Error} provides the values for $ s $ and $ t $ corresponding to the different robot orderings depicted in Fig. \ref{fig:3R-1B2D-Colinear}.
When expanded, we obtain an instance of the cubic expression \eqref{eq:Prel-Reduced-Cubic} with the coefficient $ c = - d_{12}^{\star} $ and $ d = - s \, R_{\text{bd}} $ when solving for feasible distance $ d_{12} $ while coefficient $ c = - d_{12}^{\star} $ and $ d = - t \, R_{\text{bd}} $ when we are considering distance $ d_{13} $.
Since $ d_{\text{sum}}^{\star} \in \CBR{0, \, 2} $, it follows the value for $ s $ and $ t $ can be positive or negative and hence also the coefficient $ d $ of the cubic equation. 
In turn, this may impose a condition on the desired distances $ d_{12}^{\star} $ and $ d_{13}^{\star} $ for obtaining positive values for $ d_{12} $ and $ d_{13} $ as discussed in Section \ref{subsec:Cubic-Equations}. 
In particular, we can verify that coefficient $ d $ has range $ d \in \BR{-2, \, 4} R_{\text{bd}} $. 
Taking $ d = 4 R_{\text{bd}} $, we obtain that all four robot orderings in Fig. \ref{fig:3R-1B2D-Colinear} can occur when the desired distances satisfy $ d_{ij}^{\star} \geq \sqrt{3} \sqrt[3]{2 R_{\text{bd}}} $.

In the next part, we will show that the co-linear moving formations are unstable. 

\subsection{Local stability analysis of the equilibrium and moving formations}
We have characterized the equilibrium configurations and also conditions for the moving formations. 
It is of interest to study the local stability property of these different sets. 
Similar to the stability analysis for the (\textbf{1D2B}) setup, we will use Lyapunov's indirect method. 
The Jacobian matrix corresponding to the $ z $-dynamics \eqref{eq:3R-1B2D-Link-Dynamics} results in
\begin{equation} \label{eq:3R-1B2D-Jacobian}
    \begin{aligned}
        A =
        -
        \begin{bmatrix}
            K_{\text{b}} A_{12\text{b}} + K_{\text{d}} A_{12\text{d}} & K_{\text{b}} A_{13\text{b}}
            \\
            K_{\text{b}} A_{12\text{b}} & K_{\text{b}} A_{13\text{b}} + K_{\text{d}} A_{13\text{d}}
        \end{bmatrix}
        ,
    \end{aligned}
\end{equation}
where as before, $ A_{ij\text{d}} = e_{ij\text{d}} I_{2} + 2 z_{ij} z_{ij}^{\top} $ and $ A_{ij\text{b}} = \frac{1}{d_{ij}} \BR{I_{2} - g_{ij} g_{ij}^{\top}} $, $ ij \in \CBR{12, \, 13} $.

We obtain the following result for the equilibrium configurations in \eqref{eq:3R-1B2D-Equilibrium-Sets}.

\vspace{0.5\baselineskip}
\begin{lemma} \label{lem:3R-1B2D-Jacobian-Equilibrium}
    The Jacobian matrix $ A_{\text{E}} $ at the equilibrium configurations in $ \mathcal{S}_{p}^{\text{correct}} \cup \mathcal{S}_{p}^{\text{flipped}} $ is Hurwitz. 
\end{lemma}

\vspace{0.5\baselineskip}
\begin{IEEEproof}
    For the correct and desired equilibrium configurations in $ \mathcal{S}_{p}^{\text{correct}} $ , the Jacobian matrix \eqref{eq:3R-1B2D-Jacobian} evaluates to 
    \begin{equation} \label{eq:3R-1B2D-Equilibrium-Correct-Jacobian}
        \begin{aligned}
            & A_{\text{E}}^{\text{correct}}
                = - \DIAG{
                \begin{bmatrix}
                    x^{\star} & x^{\star} & p^{\star} & p^{\star} 
                \end{bmatrix}
                }
                -
            \\
            &   
                \begin{bmatrix}
                    \BR{y^{\star} - x^{\star}} a^{2} & \BR{y^{\star} - x^{\star}} ab & p^{\star} d^{2} & -p^{\star} cd
                    \\
                    \BR{y^{\star} - x^{\star}} ab & \BR{y^{\star} - x^{\star}} b^{2} & - p^{\star}cd & p^{\star} c^{2}
                    \\
                    x^{\star} b^{2} & - x^{\star}ab & \BR{q^{\star} - p^{\star}} c^{2} & \BR{q^{\star} - p^{\star}} cd
                    \\
                    - x^{\star} ab & x^{\star} a^{2} & \BR{q^{\star} - p^{\star}} cd & \BR{q^{\star} - p^{\star}} d^{2}
                \end{bmatrix}
                ,
        \end{aligned}
    \end{equation}
    where $ x, \, y, \, p, \, q $ and the bearing matrices are previously defined in \eqref{eq:3R-1D2B-Jacobian-Variables} and \eqref{eq:3R-1D2B-Jacobian-Matrices}.
    Also, for the equilibrium configurations in $ \mathcal{S}_{p}^{\text{flipped}} $ yielding a flipped formation shape, we obtain
    \begin{equation} \label{eq:3R-1B2D-Equilibrium-Flipped-Jacobian}
        \begin{aligned}
            & A_{\text{E}}^{\text{flipped}}
                = - \DIAG{
                \begin{bmatrix}
                    x^{\star} & x^{\star} & p^{\star} & p^{\star} 
                \end{bmatrix}
                }
                -
            \\
            &   
                \begin{bmatrix}
                    \BR{y^{\star} - x^{\star}} c^{2} & \BR{y^{\star} - x^{\star}} cd & p^{\star} b^{2} & -p^{\star} ab
                    \\
                    \BR{y^{\star} - x^{\star}} cd & \BR{y^{\star} - x^{\star}} d^{2} & - p^{\star}ab & p^{\star} a^{2}
                    \\
                    x^{\star} d^{2} & - x^{\star} cd & \BR{q^{\star} - p^{\star}} a^{2} & \BR{q^{\star} - p^{\star}} ab
                    \\
                    - x^{\star} cd & x^{\star} c^{2} & \BR{q^{\star} - p^{\star}} ab & \BR{q^{\star} - p^{\star}} b^{2}
                \end{bmatrix}
                .
        \end{aligned}
    \end{equation}
    The characteristic polynomial $ \chi_{\text{E}}\BR{\lambda} $ corresponding to the Jacobian matrices $ A_{\text{E}}^{\text{correct}} $ and $ A_{\text{E}}^{\text{flipped}} $ is the same, namely
    \begin{equation} \label{eq:3R-1B2D-Equilibrium-Characteristic}
        \begin{aligned}
            \chi_{\text{E}} \BR{\lambda} 
                & = \BR{\lambda + q^{\star}} \BR{\lambda + y^{\star}} \dots 
                \\
                & \quad \BR{\lambda^{2} + \BR{p^{\star} + x^{\star}} \lambda + p^{\star} x^{\star} \sin^{2} \theta^{\star}}
                .
        \end{aligned}
    \end{equation}
    The roots of \eqref{eq:3R-1B2D-Equilibrium-Characteristic} are 
    \begin{equation} \label{eq:3R-1B2D-Equilibrium-Characteristic-Roots}
        \begin{aligned}
            \lambda_{1}
                & = -q^{\star}
                , \quad 
            \lambda_{2}
                = - y^{\star}
                ,
                \\
            \lambda_{3, \, 4}
                & = - \frac{1}{2} \BR{p^{\star} + x^{\star}} \pm \frac{1}{2} \sqrt{\BR{p^{\star} + x^{\star}}^{2} - 4 p^{\star} x^{\star} \sin^{2} \theta^{\star}}
        \end{aligned}
    \end{equation}
    We can verify that $ 0 < 4 p^{\star} x^{\star} \sin^{2} \theta^{\star} \leq \BR{p^{\star} + x^{\star}}^{2} $.
    This implies that all $ \lambda $s are real. 
    Also, $ - \BR{p^{\star} + x^{\star}} + \sqrt{\BR{p^{\star} + x^{\star}}^{2} - 4 p^{\star} x^{\star} \sin^{2} \theta^{\star}} < 0 $ and hence, we conclude that all roots are negative real. 
    This completes the proof.
\end{IEEEproof}

\vspace{0.5\baselineskip}
This leads to the following main result:

\vspace{0.5\baselineskip}
\begin{theorem}
    Consider a team of three robots arranged in the (\textbf{1B2D}) setup with closed-loop dynamics given by \eqref{eq:3R-1B2D-Closed-Loop}. 
    Given an initial configuration $ p\BR{0} $ that is close to the desired formation shape, then the robot trajectories asymptotically converge to a point $ \widehat{p} \in \mathcal{S}_{p}^{\text{correct}} $.
\end{theorem}

\vspace{0.5\baselineskip}
\begin{IEEEproof}
    Following Lemma \ref{lem:3R-1B2D-Jacobian-Equilibrium}, we obtain that link trajectories locally asymptotically converge to the desired relative positions $ z^{\star} $ when they are initialized in the neighborhood of it. 
    With $ z_{ij}^{\star} = d_{ij}^{\star} \, g_{ij}^{\star} $, it follows that also the robots converge to a point $ \widehat{p} \in \mathcal{S}_{p}^{\text{correct}} $.
\end{IEEEproof} 

\begin{table}[!tb]
    \centering
    \renewcommand{\arraystretch}{1.2}
    \caption{Values for the variables $ s $ and $ t $ corresponding to the robot orderings in Fig. \ref{fig:3R-1B2D-Colinear}}
    \label{tab:3R-1B2D-Distance-Error}
    \begin{tabularx}{0.75\linewidth}{lrr}
        \toprule 
        Ordering & $ s $ & $ t $
        \\
        \midrule
        I: $ g_{12} = g_{13} = g_{\text{sum}}^{\star} $ & $ -2 + d_{\text{sum}}^{\star} $ & $ -2 + d_{\text{sum}}^{\star} $
        \\
        II: $ g_{12} = g_{13} = - g_{\text{sum}}^{\star} $ & $ -2 - d_{\text{sum}}^{\star} $ & $ -2 - d_{\text{sum}}^{\star} $
        \\
        III: $ g_{12} = - g_{13} = g_{\text{sum}}^{\star} $ & $ d_{\text{sum}}^{\star} $ & $ - d_{\text{sum}}^{\star} $
        \\
        IV: $ g_{12} = - g_{13} = - g_{\text{sum}}^{\star} $ & $ - d_{\text{sum}}^{\star} $ & $ d_{\text{sum}}^{\star} $
        \\
        \bottomrule
    \end{tabularx}
\end{table}

\vspace{0.5\baselineskip}
Employing Lyapunov's indirect method to the moving co-linear formations yields the following statement

\vspace{0.5\baselineskip}
\begin{theorem}
    Let $ p \in \RN^{6} $ be a configuration yielding a co-linear formation satisfying conditions in Table \ref{tab:3R-1B2D-Distance-Error}. 
    Then the configuration $ p $ is unstable.
\end{theorem}

    \begin{table}[!tb]
        \centering
        \renewcommand{\arraystretch}{1.2}
        \caption{Values for the coefficients $ BB $ and $ CC $ corresponding to the robot orderings in Fig. \ref{fig:3R-1B2D-Colinear}}
        \label{tab:3R-1B2D-Characteristic-Moving-Coefficients}
        \begin{tabularx}{\linewidth}{lrr}
            \toprule 
            Ordering & $ BB $ & $ CC $
            \\
            \midrule
            I: & $ \BR{p + x} \BR{-1 + d_{\text{sum}}^{\star}} $ & $ px\BR{\BR{-1 + d_{\text{sum}}^{\star}}^{2} - 1 } < 0 $
            \\
            II: & $ \BR{p + x} \BR{-1 - d_{\text{sum}}^{\star}} < 0 $ & $ px\BR{\BR{-1 - d_{\text{sum}}^{\star}}^{2} - 1 } $
            \\
            III: & $ \BR{p + x} + \BR{x - p} d_{\text{sum}}^{\star} $ & $ - \BR{d_{\text{sum}}^{\star}}^{2} px < 0 $
            \\
            IV: & $ \BR{p + x} - \BR{x - p} d_{\text{sum}}^{\star} $ & $ - \BR{d_{\text{sum}}^{\star}}^{2} px < 0 $
            \\
            \bottomrule
        \end{tabularx}
    \end{table}
    
\vspace{0.5\baselineskip}
\begin{IEEEproof}
    We first obtain the matrix $ A_{\text{M}} $ and the corresponding characteristic polynomial $ \chi_{\text{M}} \BR{\lambda} $.
    The sub-matrices for the Jacobian matrix \eqref{eq:3R-1B2D-Jacobian} are 
    $ A_{12\text{d}} = R_{\text{bd}} \frac{s}{d_{12}} I_{2} + 2 d_{12}^{2} g_{\text{sum}}^{\star} g_{\text{sum}}^{\star \, \top} $, 
    $ A_{13\text{d}} = R_{\text{bd}} \frac{t}{d_{13}} I_{2} + 2 d_{13}^{2} g_{\text{sum}}^{\star} g_{\text{sum}}^{\star \, \top} $,
    $ A_{12\text{b}} = \frac{1}{d_{12}} \BR{I_{2} - g_{\text{sum}}^{\star} g_{\text{sum}}^{\star \, \top}} $, and 
    $ A_{13\text{b}} = \frac{1}{d_{13}} \BR{I_{2} - g_{\text{sum}}^{\star} g_{\text{sum}}^{\star \, \top}} $ with the bearing matrix $ g_{\text{sum}}^{\star}g_{\text{sum}}^{\star \, \top} = \left[\begin{smallmatrix} k^{2} & kl \\ kl & l^{2} \end{smallmatrix} \right] $.
    The values for $ s $ and $ t $ depend on the considered ordering in Table \ref{tab:3R-1B2D-Distance-Error}. 
    Hence, $ A_{\text{M}} $ takes the form
    \begin{equation} \label{eq:3R-1B2D-Moving-Jacobian}
        \begin{aligned}
            & A_{\text{M}}
            = - \BR{\DIAG{\begin{bmatrix} \BR{s + 1} x & \BR{t + 1} p \end{bmatrix}} \otimes I_{2}} - 
            \\
            & \,
            \begin{bmatrix}
                \BR{y - x} k^{2} & \BR{y - x} kl & p l^{2} &  - p k l
                \\
                \BR{y - x} kl & \BR{y - x} l^{2} & - p k l & p k^{2}
                \\
                x l^{2} & - x k l & \BR{q - p} k^{2} & \BR{q - p} kl
                \\
                - x k l & x k^{2} & \BR{q - p} kl & \BR{q - p} l^{2}
            \end{bmatrix}
            ,
        \end{aligned}
    \end{equation}
    where the variables $ x $, $ y $, $ p $, and $ q $ are defined in \eqref{eq:3R-1D2B-Jacobian-Variables}.
    The characteristic polynomial $ \chi_{\text{M}}\BR{\lambda} $ corresponding to matrix $ A_{\text{M}} $ is found to be 
    \begin{equation} \label{eq:3R-1B2D-Characteristic-Moving}
        \begin{aligned}
            \chi_{\text{M}} \BR{\lambda} 
                & = \BR{\lambda + p t + q} \BR{\lambda + x s + y} \BR{\lambda^{2} + BB \lambda + CC}
                ,
        \end{aligned}
    \end{equation}
    where the coefficients are $ BB = \BR{p\BR{t + 1} + x \BR{s + 1}} $ and $ CC = p x\BR{\BR{t + 1}\BR{s + 1} - 1} $.
    We explore the nature of the roots, hereby focusing on the coefficients of the quadratic polynomial. 
    Table \ref{tab:3R-1B2D-Characteristic-Moving-Coefficients} presents the values for $ BB $ and $ CC $ corresponding to the different robot ordering, 
    where whenever is possible, the sign of the coefficient is also provided. 
    By Lemma \ref{lem:App-Pol-Coeff-Negative}, we infer that the quadratic polynomial in \eqref{eq:3R-1B2D-Characteristic-Moving} contains at least a root with positive real part since for each ordering, either coefficient $ BB $ or $ CC $ is negative.
    This implies the matrix $ A_{\text{M}} $ is not Hurwitz; therefore, the co-linear formations are unstable.
    This completes the proof.
\end{IEEEproof}

\vspace{0.5\baselineskip}


\begin{figure*}
    \centering
    {
    \subfigure[Correct Formation]
    {
    \includegraphics[width=0.475\textwidth]{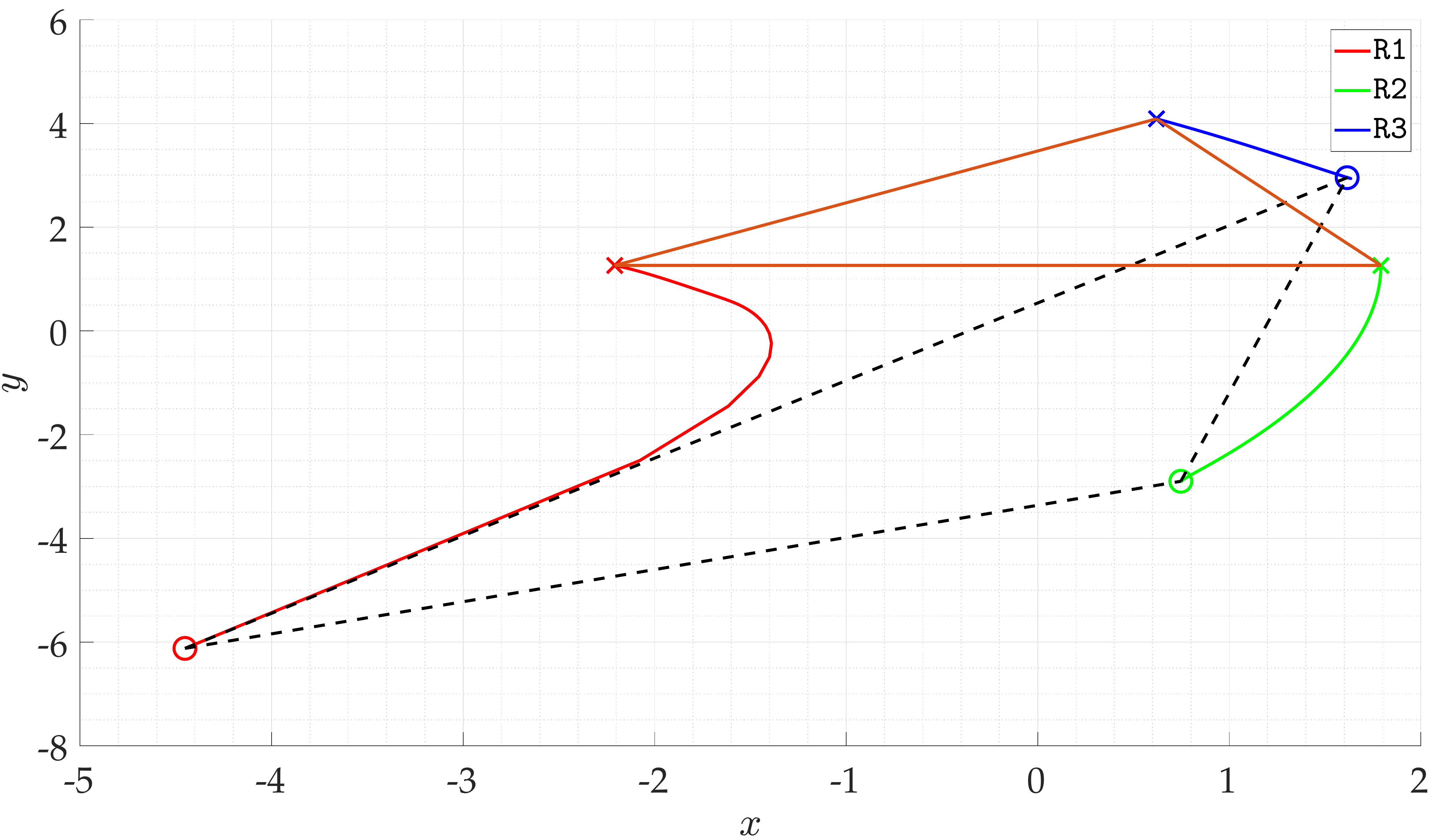} 
    \label{fig:3R-1D2B-Correct}
    }
    \hfill
    \subfigure[Moving Formation]
    {\includegraphics[width=0.475\textwidth]{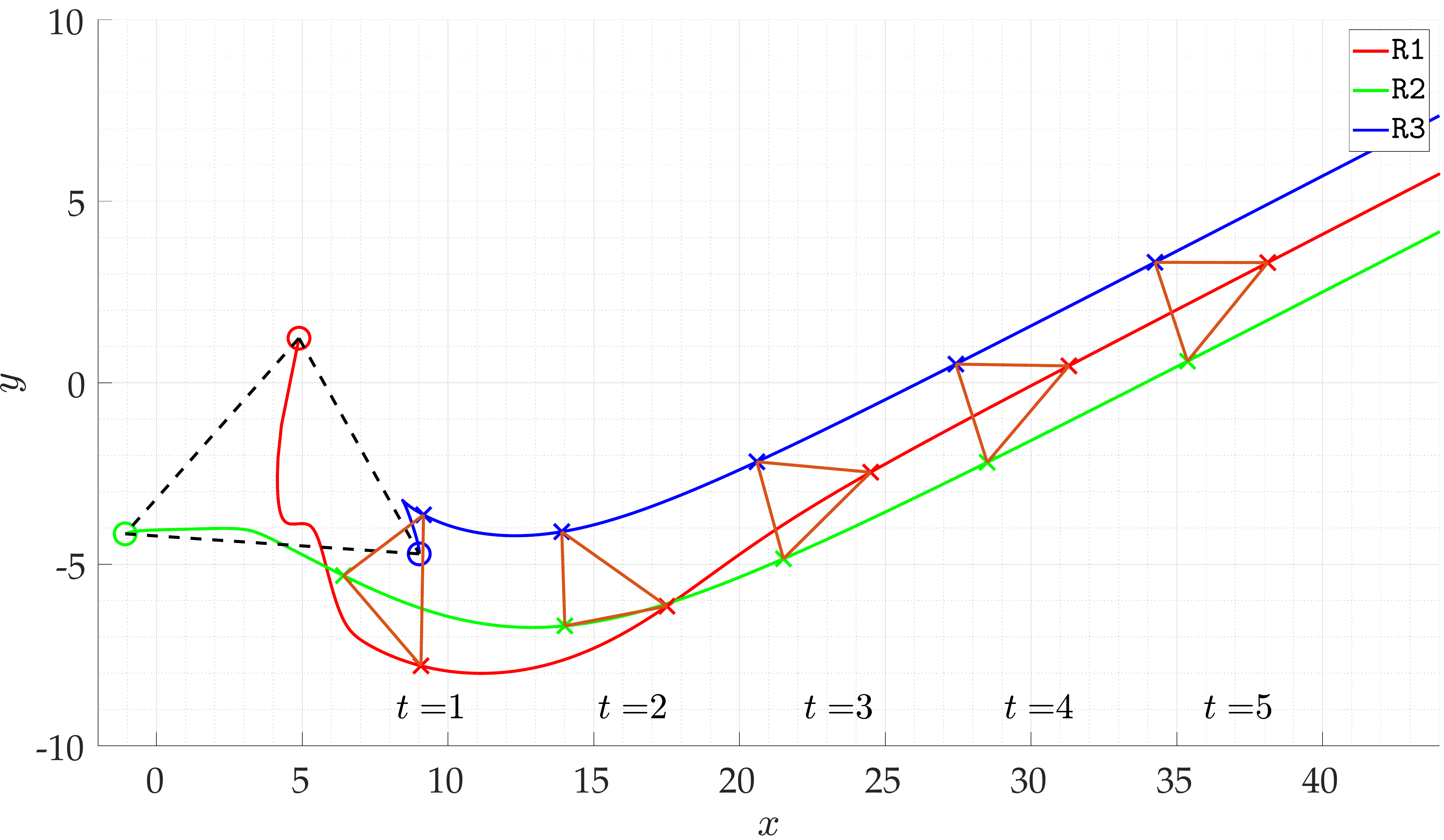}
    \label{fig:3R-1D2B-Moving}
    }
    }
    \caption{Robot trajectories for the (\textbf{1D2B}) setup; $ \circ $ represents the initial position and $ \times $ is the final position of the robot. On the left panel, we have an initial configuration (connected by dashed lines) where robots converge to the correct formation shape (connected by solid lines) while the right panel illustrates an initial configuration where robots converge to the moving configuration with velocity $ w = K_{\text{b}} b_{\text{sum}} $.}
    \label{fig:3R-1D2B-Sim-Result}
\end{figure*}

\begin{figure*}
    \centering
    {
    \subfigure[Flipped Formation]
    {
    \includegraphics[width=0.475\textwidth]{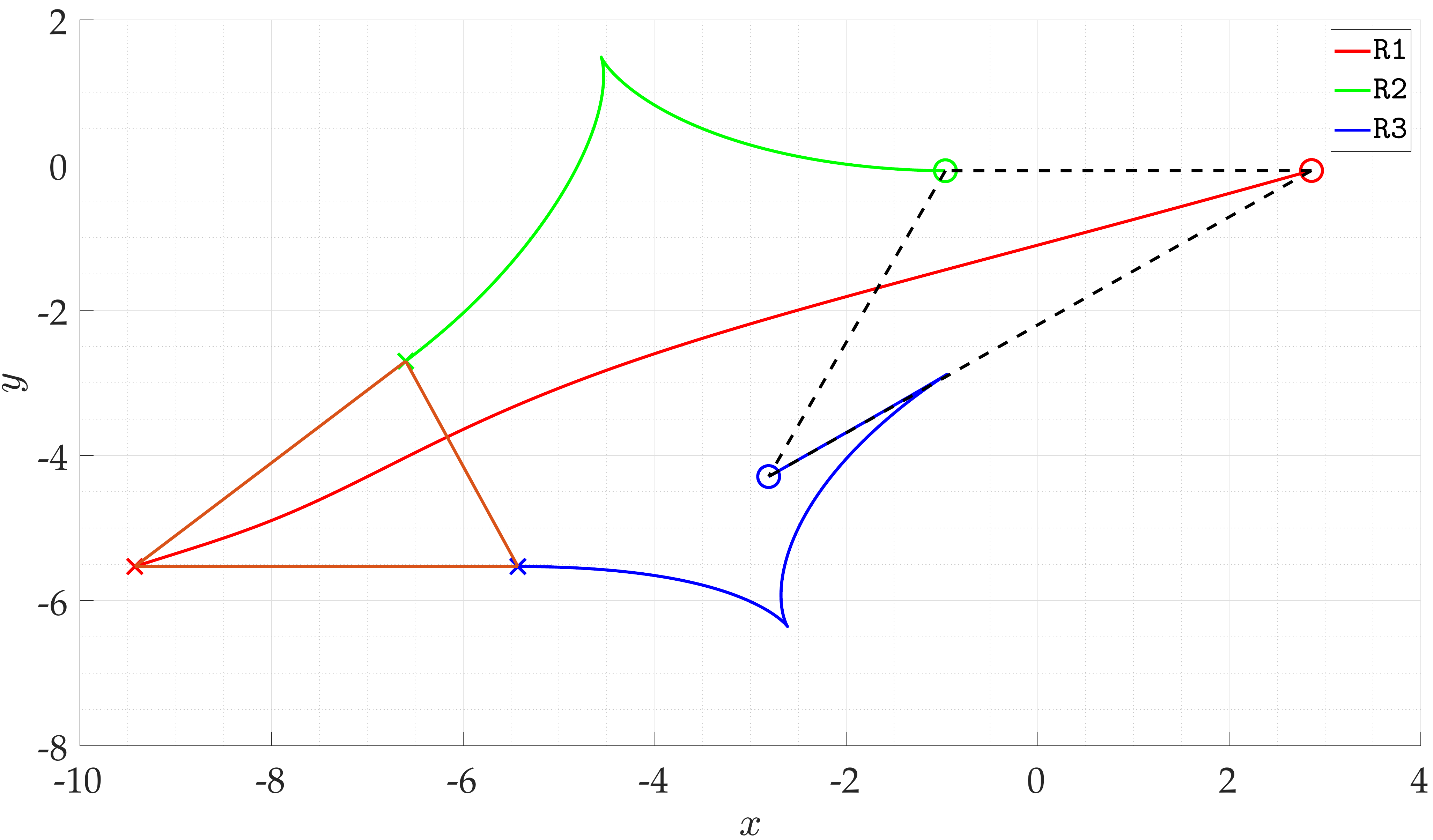} 
    \label{fig:3R-1B2D-Flipped}
    }
    \hfill
    \subfigure[Moving Formation]
    {\includegraphics[width=0.475\textwidth]{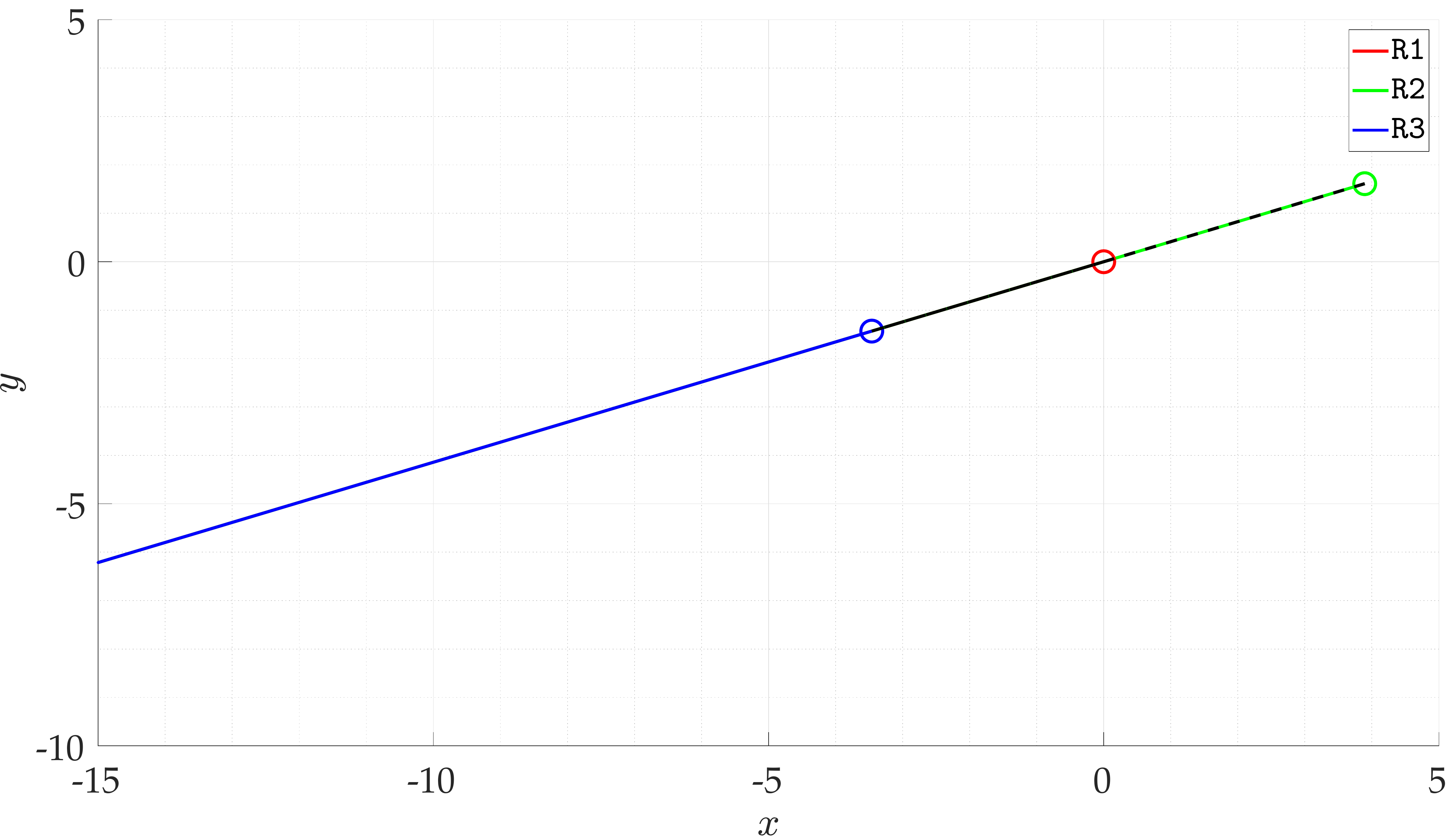} 
    \label{fig:3R-1B2D-Moving}
    }
    }
    \caption{Robot trajectories for the (\textbf{1B2D}) setup; $ \circ $ represents the initial position and $ \times $ is the final position of the robot.   On the left panel, we have an initial configuration (connected by dashed lines) where robots move and converge to the flipped formation shape while the right panel illustrates the evolution of the mobile robots when the initial configuration satisfy conditions for ordering III of the moving formation in Table \ref{tab:3R-1B2D-Characteristic-Moving-Coefficients}. The robots move with a velocity $ w = - K_{\text{b}} b_{\text{sum}} $. If the robots have slightly different initial positions, we will have convergence to an equilibrium configuration since these co-linear moving configurations are unstable.}
    \label{fig:3R-1B2D-Sim-Result}
\end{figure*}

\section{NUMERICAL EXAMPLE} \label{sec:Numerical-Example}
In this section, we present simulation results for the three robot case with interconnection structure as depicted in Fig. \ref{fig:Robot-Topology}. 
We consider two triangular formation shapes with the same distances $ d_{12}^{\star} $ and $ d_{13}^{\star} $ but different value for the internal angle $ \theta^{\star} $ (Note: $ \theta^{\star} = \cos^{-1} \BR{g_{12}^{\star \, \top} g_{13}^{\star}} $).
In particular, shape $ \mathbf{T_{1}} $ has bearing vectors such that the internal angle is $ \theta_{\mathbf{T_{1}}}^{\star} = 15^{\degree} $ while for shape $ \mathbf{T_{2}} $, we take $ \theta_{\mathbf{T_{2}}}^{\star} = 45^{\degree} $.
We set the gain ratio $ R_{\text{bd}} $ to a value $ 4 $.
Taking the different setups into consideration, the threshold distance such that moving formations (stable or unstable) exist is $ \widehat{d} = 2 \sqrt{3} \approx 3.4641 $.
We set the desired distances to $ d_{12}^{\star} = d_{13}^{\star} = 4 $ and assume $ \angle g_{12}^{\star} = 0^{\degree} $. 
Thus, shape $ \mathbf{T_{1}} $ and $ \mathbf{T_{2}} $ has the following desired constraints:
\begin{equation}
    \begin{aligned}
        \mathbf{T_{1}}:
        \, d_{12}^{\star} = d_{13}^{\star} = 4; \, \angle g_{12}^{\star} = 0^{\degree}, \, \angle g_{13}^{\star} = 15^{\degree};
        \\
        \mathbf{T_{2}}:
        \, d_{12}^{\star} = d_{13}^{\star} = 4; \, \angle g_{12}^{\star} = 0^{\degree}, \, \angle g_{13}^{\star} = 45^{\degree}
        .
    \end{aligned}
\end{equation}
For shape $ \mathbf{T_{1}} $, the moving formation for the (\textbf{1D2B}) setup is unstable, since $ \cos^{2} \BR{15^{\degree}} = 0.9330 > 0.9321 $.
Hence the constraint in \eqref{eq:3R-1D2B-Moving-Formation-Cosine} is violated.
For shape $ \mathbf{T_{2}} $, we obtain $ \cos^{2} \BR{15^{\degree}} = 0.5 < 0.9321 $ satisfying constraint \eqref{eq:3R-1D2B-Moving-Formation-Cosine}.

We remark that when desired distance constraints are provided, we can also modify the control gains $ K_{\text{d}} $ for the distance robot(s) and $ K_{\text{b}} $ for the bearing robot(s) such that the desired distances satisfy $ d_{ij}^{\star} < \widehat{d} $. 
This prevents the occurrence of (stable or unstable) moving configurations.
In the current example, we intentionally set first the gain ratio $ \textit{R}_{\text{bd}} $ and then obtain desired distances $ d_{ij}^{\star} $s in order to show the existence and local asymptotic stability of the moving formations in the (\textbf{1D2B}) setup. 

\subsection{\emph{(\textbf{1D2B})} Simulation Results}
In this part, we present simulation results for the three robots in the (\textbf{1D2B}) setup, thereby focusing on the formation shape $ \mathbf{T_{2}} $.
The Jacobian matrix $ A_{\text{M}} $ for the moving formation with distances $ d_{12} = d_{13} \approx 3.8686 $ is checked to be Hurwitz. 
Indeed, all values of the first column in the Routh-Hurwitz table \eqref{eq:3R-1D2B-RH-Criterion} evaluates to a positive value. 
Therefore, employing the closed-loop dynamics \eqref{eq:3R-1D2B-Closed-Loop} can, depending on the initial configuration $ p\BR{0} $, lead to robot trajectories moving with a constant velocity. 
In Fig. \ref{fig:3R-1D2B-Moving}, we show such an outcome for a specific initial configuration $ p\BR{0} $. 
Fig. \ref{fig:3R-1D2B-Correct} depicts an initial configuration $ p\BR{0} $ leading to convergence to the correct shape.

\subsection{\emph{(\textbf{1B2D})} Simulation Results}
In this part, we present simulation results for the three robots in the (\textbf{1B2D}) setup, thereby focusing on the formation shape $ \mathbf{T_{1}} $.
For the (\textbf{1B2D}) robot setup, there are two equilibrium formations, namely the correct and desired formation and the flipped formation satisfying only the distance constraints but not the bearing constraints.
Fig. \ref{fig:3R-1B2D-Flipped} depicts an initial configuration $ p\BR{0} $ which converges to this flipped formation.
Notice that the signed area corresponding to $ p\BR{0} $ is positive (counter-clockwise cyclic ordering of the robots) while the flipped formation has a negative signed area (clockwise cyclic ordering of the robots). 
Fig. \ref{fig:3R-1B2D-Moving} depicts an initial co-linear configuration $ p\BR{0} $ leading to the robots to move with a constant velocity when employing the closed-loop dynamics \eqref{eq:3R-1B2D-Closed-Loop}.
When perturbed, it will converge either to the correct or the flipped formation shape. 


\section{CONCLUSIONS \& FUTURE WORK} \label{sec:Conclusions}
In the current work, we have considered the formation shape problem for teams of two and three robots partitioned into two categories, namely distance and bearing robots.
Our aim is to employ gradient-based control laws in a heterogeneous setting and provide a systematic study on the stability of the possible formation shapes that arise as a result.  
We have shown that under certain conditions on the distance and bearing error signals, we obtain distorted formation shapes moving with a constant velocity $ w $. 
For the (\textbf{1D1B}) and the (\textbf{1B2D}) robot setup, these undesired formation shapes are unstable while for the (\textbf{1D2B}) robot setup, we derive conditions such that the distorted moving formation shape is locally asymptotically stable. 
Furthermore, by increasing the value for the gain ratio $ R_{\text{bd}} = \frac{K_{\text{b}}}{K_{\text{d}}} $, the occurrence of distorted moving formation shapes can be postponed. 
This may lead to global asymptotic stability of the desired formation shape, depending on the setup considered. 

We note that the moving configurations in the (\textbf{1D2B}) setup and the flipped equilibrium configuration in the (\textbf{1B2D}) setup both have a signed area that has an opposite sign compared to the signed area of the desired formation shape. 
Hence the use of signed constraints as introduced in \cite{Anderson2017, Kwon2019} is a possible future direction. 
For the (\textbf{1D2B}) setup, the inclusion of the signed area constraint in \cite{Anderson2017} does not increase the sensing load of the distance robot while it can have the potential of mitigating the existence of distorted formation shapes. 


\bibliographystyle{IEEEtran}
\bibliography{Literature}


\appendix

\subsection{Proof of Lemma \ref{lem:Prel-Reduced-Cubic-Roots-Positive}} \label{subsec:App-Lemma-Proof-1}

    First, we define $ v = -\frac{d}{2} + \sqrt{R} $.
    Since $ \Delta \geq 0 $ holds and $ R = - \frac{1}{108} \Delta $, it follows $ R \leq 0 $.
    Hence we rewrite $ v $ as the complex number $ v = -\frac{d}{2} + \sqrt{- R} \, \text{i} $.
    In polar form, we obtain $ v = r_{v} \angle \varphi_{v} $ with modulus $ r_{v} = \sqrt{- {\BR{\frac{c}{3}}}^{3}} $ and argument $ \varphi_{v} = \tan^{-1} \BR{- \frac{2}{d} \sqrt{-R}} $.
    With $ d > 0 $, we know the real part of $ v $ is negative while the imaginary part is non-negative. 
    Hence the argument $ \varphi_{v} $ is in the range $ \varphi_{v} \in \left(90^{\degree}, \, 180^{\degree} \right] $ with $ \varphi_{v} = 180^{\degree}  $ holds when $ R = 0 $.
    Furthermore, the complex conjugate of $ v $ is $ \widebar{v} = -\frac{d}{2} - \sqrt{- R} \, \text{i} = r_{v} \angle \BR{- \varphi_{v}} $.
    Substituting in \eqref{eq:Prel-Reduced-Cubic-Roots-Parameters}, we obtain $ A = \sqrt[3]{v} = \sqrt[3]{r_{v}} \angle \BR{\frac{1}{3} \varphi_{v}} $ and $ B = \sqrt[3]{\widebar{v}} = \widebar{A} $.
    Recalling $ \omega^{2} = \widebar{\omega} $, the cubic roots \eqref{eq:Prel-Reduced-Cubic-Roots} are  
    \begin{equation} \label{eq:Prel-Reduced-Cubic-Roots-Lemma}
        \begin{aligned}
            y_{1}
                = 2 \sqrt[3]{r_{v}} \cos \BR{\frac{1}{3} \varphi_{v}}
                , \:
            y_{2, \, 3}
                = 2 \sqrt[3]{r_{v}} \cos \BR{\frac{1}{3} \varphi_{v} \pm 120^{\degree}}
                .
        \end{aligned}
    \end{equation} 
    Corresponding to the range $ \varphi_{v} \in \left(90^{\degree}, \, 180^{\degree} \right] $, we have $ \frac{1}{3}\varphi_{v} \in \left(30^{\degree}, \, 60^{\degree} \right] $. 
    The positive roots are then found to be
    \begin{equation} \label{eq:Prel-Reduced-Cubic-Roots-Positive-V2}
        \begin{aligned}
            y_{1}
                = 2 \sqrt[3]{r_{v}} \cos \BR{\frac{1}{3} \varphi_{v}}
                , \:
            y_{3}
                = 2 \sqrt[3]{r_{v}} \cos \BR{\frac{1}{3} \varphi_{v} - 120^{\degree}}
        \end{aligned}
    \end{equation}
    With $ \frac{1}{3}\varphi_{v} \in \left(30^{\degree}, \, 60^{\degree} \right] $, we obtain for the range of the positive roots $ y_{1} \in \left[1, \, \sqrt{3} \right) \sqrt[3]{r_{v}} $ and $ y_{3} \in \left(0, \, 1 \right] \sqrt[3]{r_{v}} $.
    The inequality $ y_{1} \geq y_{3} $ follows and equality holds when $ \Delta = 0 \iff R = 0 $.
    From \eqref{eq:Prel-Reduced-Cubic-Roots-Parameters}, $ R = 0 $ is equivalent to $ - \BR{\frac{c}{3}}^{3} = \BR{\frac{d}{2}}^{2} $; therefore, $ r_{v} = \frac{d}{2} $.
    This completes the proof.

\subsection{Full characterization of the local stability analysis of the moving formations for the \emph{(\textbf{1D2B})} setup} \label{subsec:App-3R-1D2B-Moving-Formation}
In Lemma \ref{lem:3R-1D2B-Moving-Formation-Negative-Eigenvalue}, we considered only one of four possible combinations $ \BR{d_{12_{\text{M}}}, \, d_{13_{\text{M}}}} $ for the scenario $ d_{12}^{\star} > \widehat{d} $ and $ d_{13}^{\star} > \widehat{d} $.
Referring to Table \ref{tab:3R-1D2B-Feasible-Combinations}, we can have in total nine\footnote{In Lemma \ref{lem:3R-1D2B-Moving-Formation-Negative-Eigenvalue}, we define the variables $ m $ and $ n $. 
Since they can take values $ m \gtreqqless 0 $ and $ n \gtreqqless 0 $, we arrive also to the number nine.
} possible combinations $ \BR{d_{12_{\text{M}}}, \, d_{13_{\text{M}}}} $ when all the different scenarios are considered. 
Here, we provide the local stability analysis of the moving formations for the (\textbf{1D2B}) setup for all the remaining combinations and scenarios.

To this end, we first give the following auxiliary result that connects the sign of the coefficients of a polynomial of degree $ n $ to its roots. 

\vspace{0.5\baselineskip}
\begin{lemma} \label{lem:App-Pol-Coeff-Negative}
    Consider a \textit{polynomial} of degree $ n $
    \begin{equation}
        f\BR{y} = y^{n} + c_{1} y^{n - 1} + \cdots + c_{n}
        .
    \end{equation}
    Suppose the (distinct) roots of the equation $ f\BR{y} = 0 $ are $ \alpha_{i}, \, i = 1, \cdots, \, n $.
    Then the factored form of $ f\BR{y} $ is 
    \begin{equation}
        f\BR{y} = \BR{y - \alpha_{1}} \BR{y - \alpha_{2}} \cdots \BR{y - \alpha_{n}}
        ,
    \end{equation}
    and the sum and product of the roots $ \alpha_{i} $s are related to the coefficients $ c_{1} $ and $ c_{n} $ as $ \sum_{i = 1}^{n} \alpha_{i} = - c_{1} $ and $ \prod_{i = 1}^{n} \alpha_{i} = \BR{-1}^{n} c_{n} $, respectively.
    Furthermore, there exists at least a positive real root or a pair of complex roots with positive real part when the coefficient $ c_{1} $ is negative while an odd number of positive real roots exists when the coefficient $ c_{n} $ is negative. 
\end{lemma}

\vspace{0.5\baselineskip}
The proof of Lemma \ref{lem:App-Pol-Coeff-Negative} is straightforward and therefore omitted.

\vspace{0.5\baselineskip}
\begin{proposition} \label{prop:App-Expression-h}
    Consider the desired distance $ d_{ij}^{\star} \geq \widehat{d} $ with $ \widehat{d} = \sqrt{3} \sqrt[3]{\frac{R_{\text{bd}}}{2}} $ and let $ d_{ij} $ be a feasible solution to the cubic equation \eqref{eq:3R-1D2B-Cubic-equation}. 
    Define the variable $ h\BR{d_{ij}} = 2 K_{\text{d}} d_{ij}^{2} - K_{\text{b}} \frac{1}{d_{ij}} $.
    Then,
    \begin{itemize}
        \item 
        $ h\BR{d_{ij}} = 0 $ when $ d_{ij}^{\star} = \widehat{d} $;
        
        \item 
        $ h\BR{d_{ij}} > 0 $ when $ d_{ij}^{\star} > \widehat{d} \wedge d_{ij} = y_{\text{p}_{1}}\BR{d_{ij}^{\star}} $ in Lemma \ref{lem:Prel-Reduced-Cubic-Roots-Positive};
        
        \item 
        $ h\BR{d_{ij}} \gtreqqless 0 $ when $ d_{ij}^{\star} > \widehat{d} \wedge d_{ij} = y_{\text{p}_{2}}\BR{d_{ij}^{\star}} $ in Lemma \ref{lem:Prel-Reduced-Cubic-Roots-Positive}.
    \end{itemize}
\end{proposition}

\vspace{0.5\baselineskip}
\begin{IEEEproof}
    The expression $ h\BR{d_{ij}} \gtreqqless 0 $ is equivalent to $ d_{ij} \gtreqqless \sqrt[3]{\frac{R_{\text{bd}}}{2}} = \frac{1}{3} \sqrt{3} \widehat{d} $.
    When $ d_{ij}^{\star} = \widehat{d} $, we obtain that $ d_{ij} = \frac{1}{3} \sqrt{3} \widehat{d} $, resulting in the term $ h\BR{d_{ij}} = 0 $.
    For $ d_{ij}^{\star} > \widehat{d} $, Lemma \ref{lem:Prel-Reduced-Cubic-Roots-Positive} provides two feasible distances satisfying the cubic equation \eqref{eq:3R-1D2B-Cubic-equation}. 
    The ranges of the two feasible distances are $ y_{\text{p}_{1}} \in \BR{\frac{1}{3} \sqrt{3}, \, 1} d_{ij}^{\star} $ and $ y_{\text{p}_{2}} \in \BR{0, \, \frac{1}{3} \sqrt{3}} d_{ij}^{\star} $.
    Since $ d_{ij}^{\star} > \widehat{d} $, we obtain $ y_{\text{p}_{1}} > \frac{1}{3} \sqrt{3} d_{ij}^{\star} > \frac{1}{3} \sqrt{3} \widehat{d} $ and therefore, $ h \BR{y_{\text{p}_{1}}} > 0 $. 
    For the feasible distance $ y_{\text{p}_{2}} $, we have $ y_{\text{p}_{2}} < \frac{1}{3} \sqrt{3} d_{ij}^{\star} $. 
    It also satisfies $ y_{\text{p}_{2}} \gtreqqless \frac{1}{3} \sqrt{3} \widehat{d} $ and consequently, $ h \BR{y_{\text{p}_{2}}} \gtreqqless 0 $.
    This completes the proof.
\end{IEEEproof}

\begin{table}[!tb]
    \centering
    \caption{Sign of the coefficients corresponding to the characteristic polynomial $ \chi_{\text{M}} \BR{\lambda} $ in \eqref{eq:3R-1D2B-Moving-Characteristic-Coefficients} for $ 0 < \sin^{2} \theta^{\star} < 1 $; The $ \mbox{?`} $ symbol means that the sign is indeterminate.}
    \label{tab:App-Moving-Formation-Characteristic-Coefficients-sin<0}
    \begin{tabularx}{0.675\linewidth}{lcccc}
        \toprule 
        & $ c_{1} $ & $ c_{2} $ & $ c_{3} $ & $ c_{4} $
        \\
        \midrule
        $ m < 0 \wedge n < 0 $ & $ < 0 $ & $ < 0 $ & $ > 0 $ & $ > 0 $
        \\
        $ m < 0 \wedge n = 0 $ & $ < 0 $ & $ < 0 $ & $ > 0 $ & $ = 0 $
        \\
        $ m < 0 \wedge n > 0 $ & $ \mbox{?`} $ & $ \mbox{?`} $ & $ \mbox{?`} $ & $ < 0 $
        \\
        $ m = 0 \wedge n < 0 $ & $ < 0 $ & $ < 0 $ & $ > 0 $ & $ = 0 $
        \\
        $ m = 0 \wedge n = 0 $ & $ = 0 $ & $ < 0 $ & $ = 0 $ & $ = 0 $
        \\
        $ m = 0 \wedge n > 0 $ & $ > 0 $ & $ \mbox{?`} $ & $ < 0 $ & $ = 0 $
        \\
        $ m > 0 \wedge n < 0 $ & $ \mbox{?`} $ & $ \mbox{?`} $ & $ \mbox{?`} $ & $ < 0 $
        \\
        $ m > 0 \wedge n = 0 $ & $ > 0 $ & $ \mbox{?`} $ & $ < 0 $ & $ = 0 $
        \\
        \bottomrule
    \end{tabularx}
\end{table}

\begin{table}[!tb]
    \centering
    \caption{Sign of the coefficients corresponding to the characteristic polynomial $ \chi_{\text{M}} \BR{\lambda} $ in \eqref{eq:3R-1D2B-Moving-Characteristic-Coefficients} for $ \sin^{2} \theta^{\star} = 1 $; The $ \mbox{?`} $ symbol means that the sign is indeterminate. }
    \label{tab:App-Moving-Formation-Characteristic-Coefficients-sin=0}
    \begin{tabularx}{0.675\linewidth}{lcccc}
        \toprule 
        & $ c_{1} $ & $ c_{2} $ & $ c_{3} $ & $ c_{4} $
        \\
        \midrule
        $ m < 0 \wedge n < 0 $ & $ < 0 $ & $ < 0 $ & $ > 0 $ & $ > 0 $
        \\
        $ m < 0 \wedge n = 0 $ & $ < 0 $ & $ < 0 $ & $ = 0 $ & $ = 0 $
        \\
        $ m < 0 \wedge n > 0 $ & $ \mbox{?`} $ & $ \mbox{?`} $ & $ < 0 $ & $ < 0 $
        \\
        $ m = 0 \wedge n < 0 $ & $ < 0 $ & $ < 0 $ & $ = 0 $ & $ = 0 $
        \\
        $ m = 0 \wedge n = 0 $ & $ = 0 $ & $ = 0 $ & $ = 0 $ & $ = 0 $
        \\
        $ m = 0 \wedge n > 0 $ & $ > 0 $ & $ > 0 $ & $ = 0 $ & $ = 0 $
        \\
        $ m > 0 \wedge n < 0 $ & $ \mbox{?`} $ & $ \mbox{?`} $ & $ < 0 $ & $ < 0 $
        \\
        $ m > 0 \wedge n = 0 $ & $ > 0 $ & $ > 0 $ & $ = 0 $ & $ = 0 $
        \\
        \bottomrule
    \end{tabularx}
\end{table}

\vspace{0.5\baselineskip}
Following Proposition \ref{prop:App-Expression-h}, we define the variables $ m \coloneqq h\BR{d_{12}} = y - x $ and $ n \coloneqq h\BR{d_{13}} = q - p $ in 
the characteristic polynomial $ \chi_{\text{M}} \BR{\lambda} $ \eqref{eq:3R-1D2B-Moving-Characteristic}. 
The case $ m > 0 $ and $ n > 0 $ has already been considered in Lemma \ref{lem:3R-1D2B-Moving-Formation-Negative-Eigenvalue}. 

We are ready to consider the remaining combinations of $ m $ and $ n $.
First, we assume the bearing vectors $ g_{12}^{\star} $ and $ g_{13}^{\star} $ are not perpendicular. 
This is equivalent to $ 0 < \sin^{2} \theta^{\star} < 1 $.
Table \ref{tab:App-Moving-Formation-Characteristic-Coefficients-sin<0} provides the sign of the coefficients of $ \chi_{\text{M}} \BR{\lambda} $ in \eqref{eq:3R-1D2B-Moving-Characteristic-Coefficients}.
Applying Lemma \ref{lem:App-Pol-Coeff-Negative}, we obtain that $ \chi_{\text{M}} \BR{\lambda} $ contains at least a root with positive real part for each combination of $ m $ and $ n $ in Table \ref{tab:App-Moving-Formation-Characteristic-Coefficients-sin<0}. 
This in turn implies the Jacobian matrix $ A_{\text{M}} $ in \eqref{eq:3R-1D2B-Moving-Jacobian} is not Hurwitz and hence the moving formations are unstable.

We proceed with the case in which $ \sin^{2} \theta^{\star} = 1 $, i.e., the bearing vectors satisfy $ g_{12}^{\star} \perp g_{13}^{\star} $. 
Again, we determine the sign of the coefficients for each combination of $ m $ and $ n $; see Table \ref{tab:App-Moving-Formation-Characteristic-Coefficients-sin=0}.
Applying Lemma \ref{lem:App-Pol-Coeff-Negative}, we obtain that $ \chi_{\text{M}} \BR{\lambda} $ contains at least a root with positive real part for all combination of $ m $ and $ n $ in Table \ref{tab:App-Moving-Formation-Characteristic-Coefficients-sin=0}, except for the cases $ \BR{m, \, n} = \CBR{\BR{0, \, 0}, \, \BR{> 0, \, 0}, \, \BR{0, \, > 0}} $. 
Further investigation reveals that for the case $ \BR{m, \, n} = \BR{0, \, 0} $, we obtain the root $ \lambda = 0 $ with multiplicity $ 4 $ while for $ \BR{m, \, n} = \CBR{\BR{> 0, \, 0}, \, \BR{0, \, > 0}} $, the roots to $ \chi_{\text{M}}\BR{\lambda} $ are $ \lambda_{1, \, 2} = 0 $ and $ \lambda_{3, \, 4} = \frac{1}{2} \BR{-c_{1} \pm \sqrt{c_{1}^{2} - 4 c_{2}}} $. 
Since both $ c_{1} > 0 $ and $ c_{2} > 0 $, we obtain $ \sqrt{c_{1}^{2} - 4 c_{2}} < c_{1} $ and hence $ \lambda_{3, \, 4} $ are negative real roots or are roots containing a negative real part. 
For these specific cases, the Jacobian matrix $ A_{\text{M}} $ in \eqref{eq:3R-1D2B-Moving-Jacobian} cannot provide any conclusion on the local stability while for the cases containing a root with positive real part, we conclude those moving formations are unstable. 
It should be remarked that the case $ \BR{m, \, n} = \BR{0, \, 0} $ corresponds to the scenario when $ d_{12}^{\star} = d_{13}^{\star} = \widehat{d} $, $ \BR{m, \, n} = \BR{> 0, \, 0} $ corresponds to $ d_{12}^{\star} > \widehat{d} $ and $ d_{13}^{\star} = \widehat{d} $, and $ \BR{m, \, n} = \BR{0, \, > 0} $ corresponds to $ d_{12}^{\star} = \widehat{d} $ and $ d_{13}^{\star} > \widehat{d} $.
In particular, the scenario of $ d_{12}^{\star} = d_{13}^{\star} = \widehat{d} $ is very specific. 


\begin{IEEEbiography}[{\includegraphics[width=1in, height=1.25in, clip, keepaspectratio]{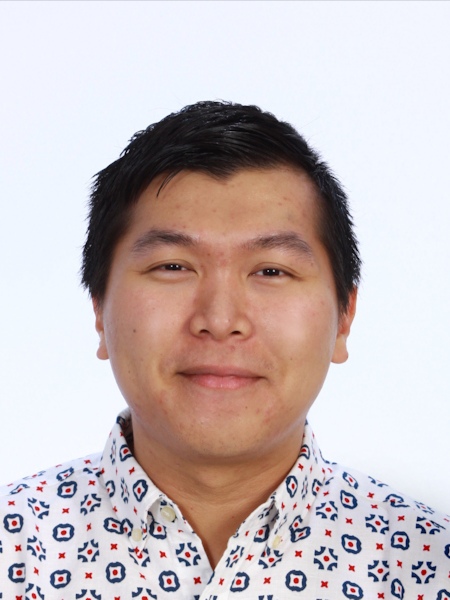}}]{Nelson Chan}
(S'18) received the B.Sc. degree in mechanical engineering from the Anton de Kom University of Suriname, Paramaribo, Suriname, in 2013, and the M.Sc. degree in systems and control from the University of Twente, Enschede, The Netherlands, in 2016. 
He is currently working toward the Ph.D. degree at the University of Groningen, Groningen, The Netherlands, under the supervision of Prof. Jayawardhana and Prof. Scherpen. 
His research interests include control and analysis of multi-agent systems. 
\end{IEEEbiography}

\begin{IEEEbiography}[{\includegraphics[width=1in, height=1.25in, clip, keepaspectratio]{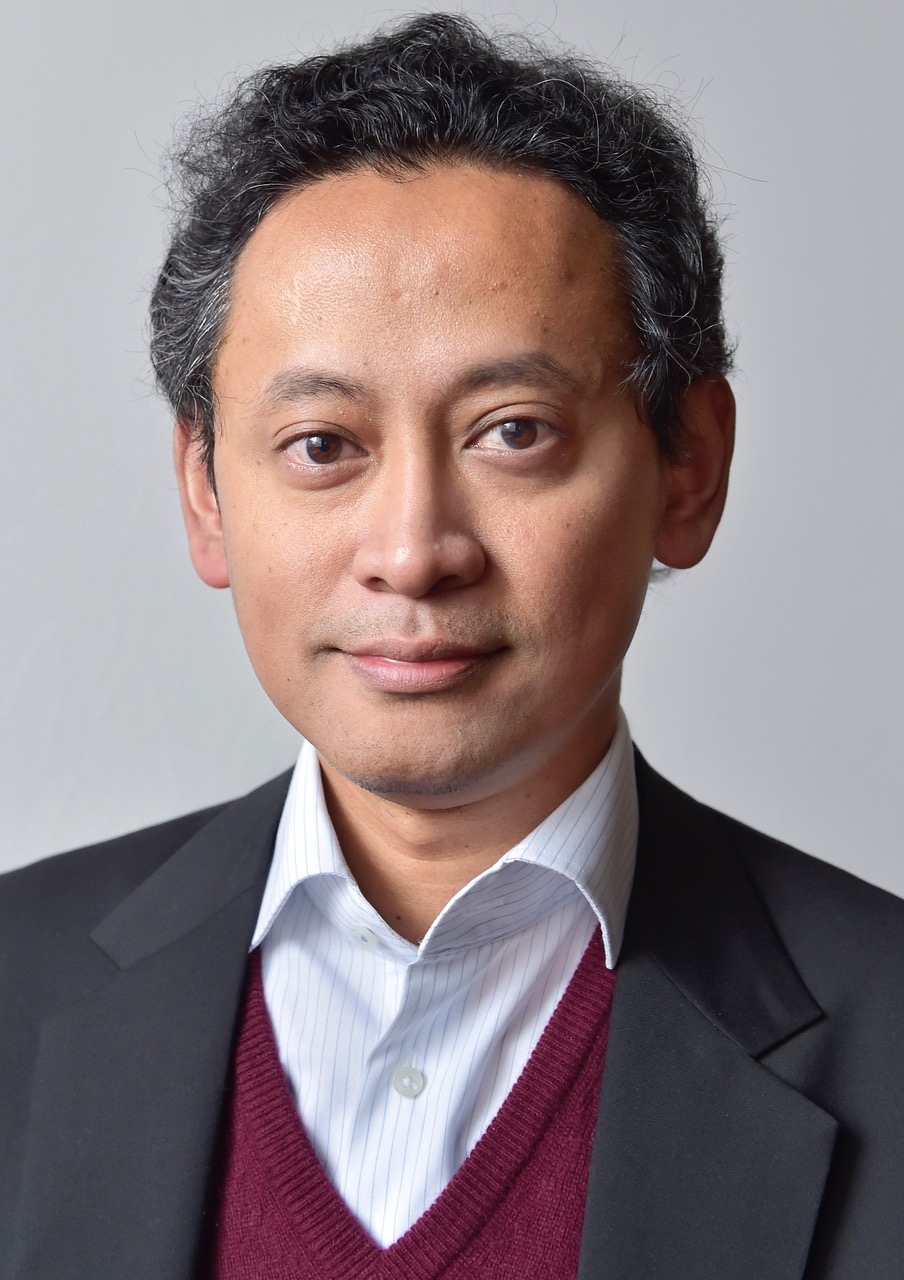}}]{Bayu Jayawardhana}
(SM’13) received the B.Sc. degree in electrical and electronics engineering from the Institut Teknologi Bandung, Bandung, Indonesia, in 2000, the M.Eng. degree in electrical and electronics engineering from the Nanyang Technological University, Singapore, in 2003, and the Ph.D. degree in electrical and electronics engineering from Imperial College London, London, U.K., in 2006. 
He is currently a Full Professor in the Faculty of Science and Engineering, University of Groningen, Groningen, The Netherlands. 
He was with Bath University, Bath, U.K., and with Manchester Interdisciplinary Bio-centre, University of Manchester, Manchester, U.K. 
His research interests include the analysis of nonlinear systems, systems with hysteresis, mechatronics, systems and synthetic biology. 
Prof. Jayawardhana is a Subject Editor of the International Journal of Robust and Nonlinear Control, an Associate Editor of the European Journal of Control and of IEEE Trans. Control Systems Technology and a member of the Conference Editorial Board of the IEEE Control Systems Society.
\end{IEEEbiography}

\begin{IEEEbiography}[{\includegraphics[width=1in, height=1.25in, clip, keepaspectratio]{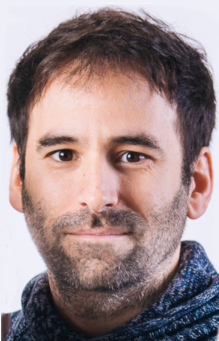}}]{H\'{e}ctor Garcia de Marina}
(M’16) received the Ph.D. degree in systems and control from the University of Groningen, the Netherlands, in 2016. He was a post-doctoral research associate with the Ecole Nationale de l'Aviation Civile, Toulouse, France, and an assistant professor at the Unmanned Aerial Systems Center at the University of Southern Denmark. He is currently a researcher in the Department of Computer Architecture and Automatic Control at the Faculty of Physics, Universidad Complutense de Madrid, Spain.
His current research interests include multi-agent systems and the design of guidance, navigation, and control systems for autonomous vehicles.
\end{IEEEbiography}


\end{document}